\documentclass[preprint,showkeys,secnumarabic,amsfonts,showpacs,amsmath,amssymb]{revtex4}
\usepackage[dvips]{color}
\usepackage{epsfig}
\usepackage{amsmath}
\usepackage{graphicx}
\def\Box{\hbox{$\rlap{$\sqcup$}\sqcap$}}

\begin{document}

\title{\bf Attractors, Statefinders and Observational Measurement for Chameleonic Brans--Dicke Cosmology }
\author{Hossein Farajollahi}
\email{hosseinf@guilan.ac.ir}

\author{Amin Salehi}
\email{a.salehi@guilan.ac.ir}
\affiliation{Department of Physics, University of Guilan, Rasht, Iran}

\date{\today}

\begin{abstract}
We investigate chameleonic Brans--Dicke model
applied to the FRW universes. A framework to study
stability and attractor solutions in the phase space is developed for the model.
We show that depending on the matter field and stability conditions,
it is possible to realize phantom-like behavior without introducing phantom filed in the model
 while the stability is fulfilled and phantom crossing occurs.
The statefinder parameters to the model for different kinds of matter interacting with
the chameleon scalar field are studied. We also compare our model with present day observations.

\end{abstract}



\maketitle

\section{Introduction}

Recently, the observations of high redshift type Ia
supernovae, the surveys of clusters of galaxies \cite{Reiss}--\cite{Riess2}, Sloan digital sky survey (
SDSS)~\cite{Abazajian} and Chandra X--ray observatory~\cite{Allen} reveal the universe accelerating expansion
and that the density of matter is very much less than the critical density. Also the
observations of Cosmic Microwave Background (CMB)
anisotropies \cite{Bennett} indicate that the universe is flat and the total energy
density is very close to the critical one \cite{Spergel}. The observations though determines basic cosmological parameters
with high precisions and strongly indicates that the universe
presently is dominated by a smoothly distributed and slowly
varying dark energy (DE) component, but at the same time they poses a serious
problem about the origin of DE \cite{Tsujikawa}. A dynamical equation of state ( EoS) parameter that is connected directly to the evolution of the energy density in the universe and indirectly to the expansion of the Universe can be regarded as a suitable parameter to explain the acceleration and the origin of DE \cite{Seljak}--\cite{Setare}. In scalar-tensor theories \cite{Sahoo}--\cite{Nojiri3}, interaction of the scalar field with matter ( for example in chameleon cosmology) \cite{Setare1}--\cite{Dimopoulos} or with geometry in Brans-Dicke (BD) cosmological models \cite{Damouri}--\cite{Biswass} can be used to interpret the late time acceleration.

On the other hand, by using the well-known geometric variables, Hubble parameter and deceleration parameter together with the new geometrical variables, the cosmological
diagnostic pair $\{r, s\}$ ( or statefinder parameters)\cite{Sahni}, the acceleration expansion of the universe and differentiation among the cosmological models can be explained in order to better fit the observational data. Moreover, since the cosmic acceleration affects the expansion history of the
universe, to understand the true nature of the driving force,
mapping of the cosmic expansion of the universe is very
crucial~\cite{Linder}. Hence, one requires various observational
probes in different redshift ranges to understand the expansion
history of the universe. One of these tests is the difference in distance modulus measurement of
 type Ia supernovae that helps us to testify the cosmological models.

In here, in addition to the most of the expectations presented in~\cite{Khourym}--\cite{Brax2}, we investigate the integration of both BD and Chameleon theory, which we call it "Chameleonic Brans--Dicke" ( CBD) model, to describe the late time acceleration of the universe and examine the dynamic of the universe, by applying stability analysis qualitatively. The non–minimal coupling term in the BD theory is equivalent to the presence
of conformal transformations such that one can change the form of the coupling term by
applying these transformations. Among these transformations, the ones which can transform
the coupling term into a constant are more attractive, for they can be employed to switch
between the Jordan and Einstein frames, which are the most discussed conformal frames
~\cite{faraon1999}--\cite{martin2010}. Nonetheless, the physics is not invariant under conformal
transformations, except in the weak gravitational field limit, and one should choose one of
them as the physical frame. As the physical frame labels the frame we live in, it can be
selected upon physical grounds consistent with principles and observations. In here, the CBD model in Jordan frame allows scalar field that is very light on cosmological
scales, to couple to matter much more strongly than gravity does,
also to couple to gravity and yet still satisfies the current
experimental and observational constraints. The cosmological value
of such a field evolves over Hubble time-scales and could
potentially cause the late--time acceleration of our universe. From a chameleonic point of view, since the field mimics the background radiation/matter field, subdominant for most of the evolution history except at late times  when it becomes dominant, it may be regarded as a cosmological tracker field \cite{farajollahi}. From a BD cosmological point of view the scalar field is coupled to the geometry and may be regarded as an intermediate field to connect the field to the geometric variables in the model. In the BD model, the weak energy condition (WEC) imposes constraint whose validity is studied in the light of its ability to explain the observational data where, in particular, for an accelerating universe and dynamical EoS parameter the WEC is satisfied \cite{Boisseau}. In CBD model where a direct coupling
 between the scalar field and the Lagrangian of the matter field is also included, the subject needs further studies and analysis.

In this paper, we study the detailed evolution of the scalar field in CBD model and the attractor property of its solutions. Section two is devoted to a detailed formulation of the cosmological model. In Section three, we obtain the autonomous equations of the model and by using the phase plane analysis qualitatively investigate the dynamic of the system and the existence of a late time attractor solution. In Section four, we examine the behavior of the EoS parameter and deceleration parameter of the model and also perform a statefinder diagnostic for the model and analyze the evolving trajectories of the model in the statefinder parameter plane. To verify our model we also compare the distance modulus measurements versus redshift z derived in the model with the data obtained from the observations of type Ia supernovae. In Section five, we present summary and conclusion.

\section{The model}

We consider the CBD gravity in the Jordan frame and in the presence of matter with the action given by,
\begin{eqnarray}\label{ac}
S=\int d^{4}x\frac{\sqrt{-g}}{2}[\phi
R-\frac{\omega}{\phi}\phi^{,\alpha}\phi_{,\alpha}-2V(\phi)+2L_{m}f(\phi)],
\end{eqnarray}
where $R$ is the Ricci scalar, $\phi$ is the Chameleon and $/$ or BD scalar field with the potential $V(\phi)$, and
$\omega$ is $BD$ parameter. We assume that $f(\phi)=f_{0}\phi^\kappa$ and $V(\phi)=V_{0}\phi^\delta$ where $\kappa$ and $ \delta $ are  dimensionless constants characterizing the slope of potential $V(\phi)$ and $f(\phi)$.
Unlike the usual Einstein--Hilbert action, the matter Lagrangian
$L_{m}$ is modified as $f(\phi)L_{m}$, where $f(\phi)$ is an
analytic function of the scalar field. The last term in the action
brings about the nonminimal interaction between
matter and the scalar field. Variation of the action (\ref{ac})
with respect to the metric tensor $g_{\mu\nu}$ gives,
\begin{eqnarray}\label{fieldeq}
G_{\mu\nu}=\frac{\omega}{\phi^{2}}[\phi_{,\mu}\phi_{,\nu}-\frac{1}{2}g_{\mu\nu}\phi_{,\alpha}\phi^{,\alpha}]
+\frac{1}{\phi}[\phi_{,\mu;\nu}-g_{\mu\nu}\Box \phi]-\frac{V(\phi)}{\phi}g_{\mu\nu}+\frac{f(\phi)}{\phi}T_{\mu\nu}.
\end{eqnarray}
In FRW cosmology, the field equation (\ref{fieldeq}) becomes,
\begin{eqnarray}\label{fried1}
3H^2=\frac{\rho_{m}f(\phi)}{\phi}-3H\frac{\dot{\phi}}{\phi}
+\frac{\omega}{2}\frac{\dot{\phi}^{2}}{\phi^{2}}-\frac{V(\phi)}{\phi},
\end{eqnarray}
\begin{eqnarray}\label{fried2}
2\dot{H}+3H^2=-\frac{p_{m}f(\phi)}{\phi}-2H\frac{\dot{\phi}}{\phi}
-\frac{\omega}{2}\frac{\dot{\phi}^{2}}{\phi^{2}}-\frac{\ddot{\phi}}{\phi}-\frac{V(\phi)}{\phi}.
\end{eqnarray}
On the other hand, the variation of the action (\ref{ac}) with respect to the spatially homogeneous scalar fields $\phi(t)$ gives,
\begin{eqnarray}\label{phiequation}
\ddot{\phi}+3H\dot{\phi}=\frac{(\rho_{m}-3p_{m})f(\phi)}{3+2\omega}
-\frac{2(2V(\phi)-\phi V')}{3+2\omega}
-\frac{(\rho_{m}-3p_{m})\phi f'}{2(3+2\omega)},
\end{eqnarray}
where prime means derivative with respect to $\phi$. We assume that the universe is filled with the barotropic fluid
with the EoS to be $p_{m}=\gamma\rho_{m}$. From equations (\ref{fried1}), (\ref{fried2}) and (\ref{phiequation}), one can easily arrive at the modified conservation equation,
\begin{eqnarray}
\dot{(\rho_{m}f)}+3\frac{\dot{a}}{a}(1+\gamma)\rho_{m}f=\frac{1}{4}(1-3\gamma)\rho_{m}\dot{f},
\end{eqnarray}
 which readily integrates to yield,
\begin{eqnarray}
\rho_{m}=\frac{M}{f^{\frac{3}{4}(1+\gamma)}a^{3(1+\gamma)}},
\end{eqnarray}
where $M$  is a constant of integration. From equations (\ref{fried1})
and (\ref{fried2}) and in comparison
with the standard Friedmann equations, we identify the effective energy density, $\rho_{eff}$, and effective pressure, $p_{eff}$, for the model as:
\begin{eqnarray}\label{rhoeff}
\rho_{eff}\equiv\frac{\rho_{m}f(\phi)}{\phi}-3H\frac{\dot{\phi}}{\phi}
+\frac{\omega}{2}\frac{\dot{\phi}^{2}}{\phi^{2}}-\frac{V(\phi)}{\phi},
\end{eqnarray}
\begin{eqnarray}\label{peff}
p_{eff}\equiv\frac{p_{m}f(\phi)}{\phi}+2H\frac{\dot{\phi}}{\phi}
+\frac{\omega}{2}\frac{\dot{\phi}^{2}}{\phi^{2}}+\frac{\ddot{\phi}}{\phi}+\frac{V(\phi)}{\phi}.
\end{eqnarray}
Now by using equations (\ref{rhoeff})
and (\ref{peff}) the conservation equation can be obtained as,
\begin{eqnarray}\label{rhodot}
\dot{\rho}_{eff}+3H(1+\omega_{eff})\rho_{eff}=0,
\end{eqnarray}
with an effective EoS, $p_{eff}=\omega_{eff}\rho_{eff}$. The numerical calculation of the dynamical effective EoS parameter $\omega_{eff}$ and deceleration parameter $q$ for the model are shown in  Fig 1):\\

\begin{tabular*}{2.5 cm}{cc}
\includegraphics[scale=.35]{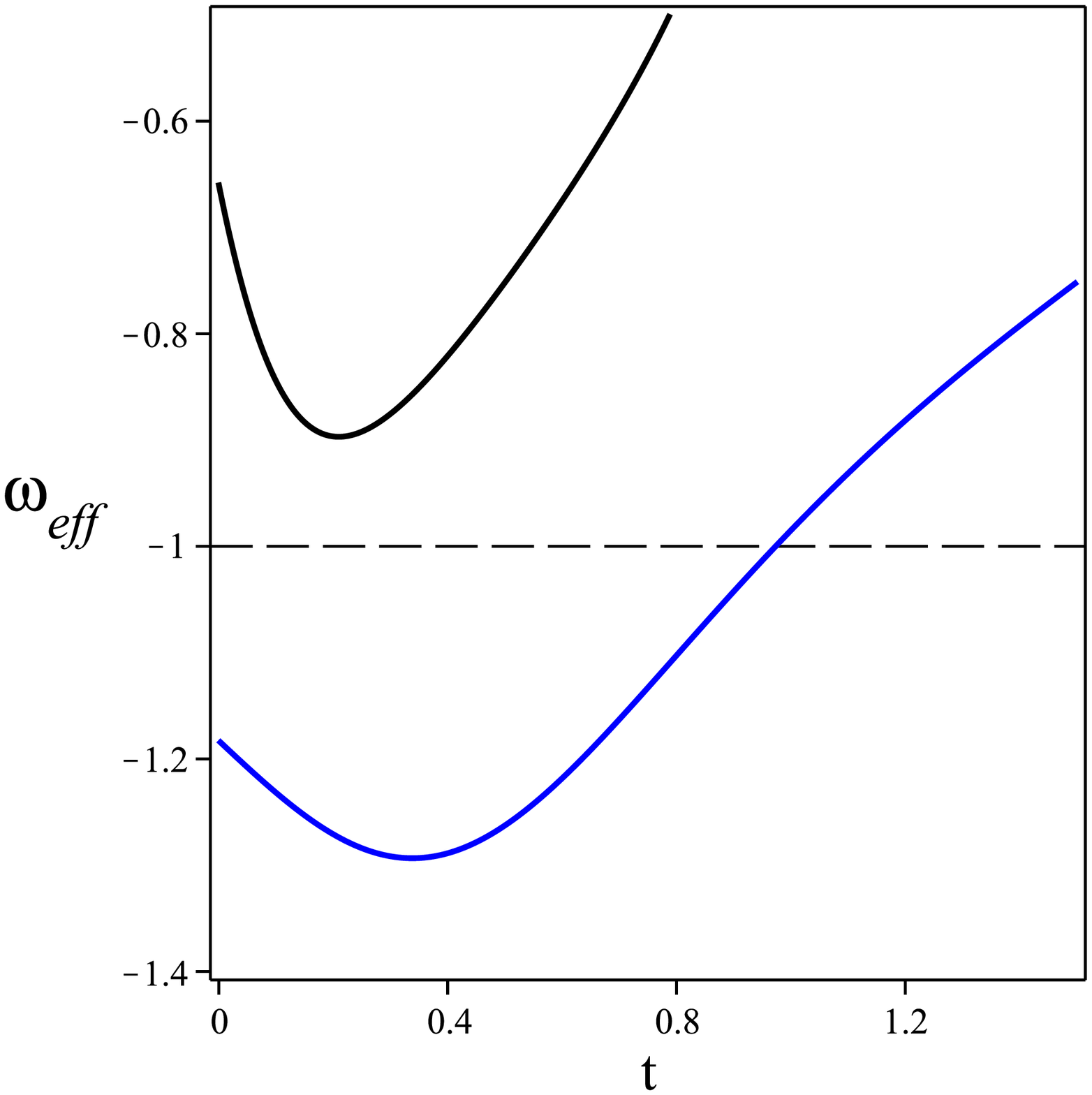}\hspace{0.1 cm}\includegraphics[scale=.35]{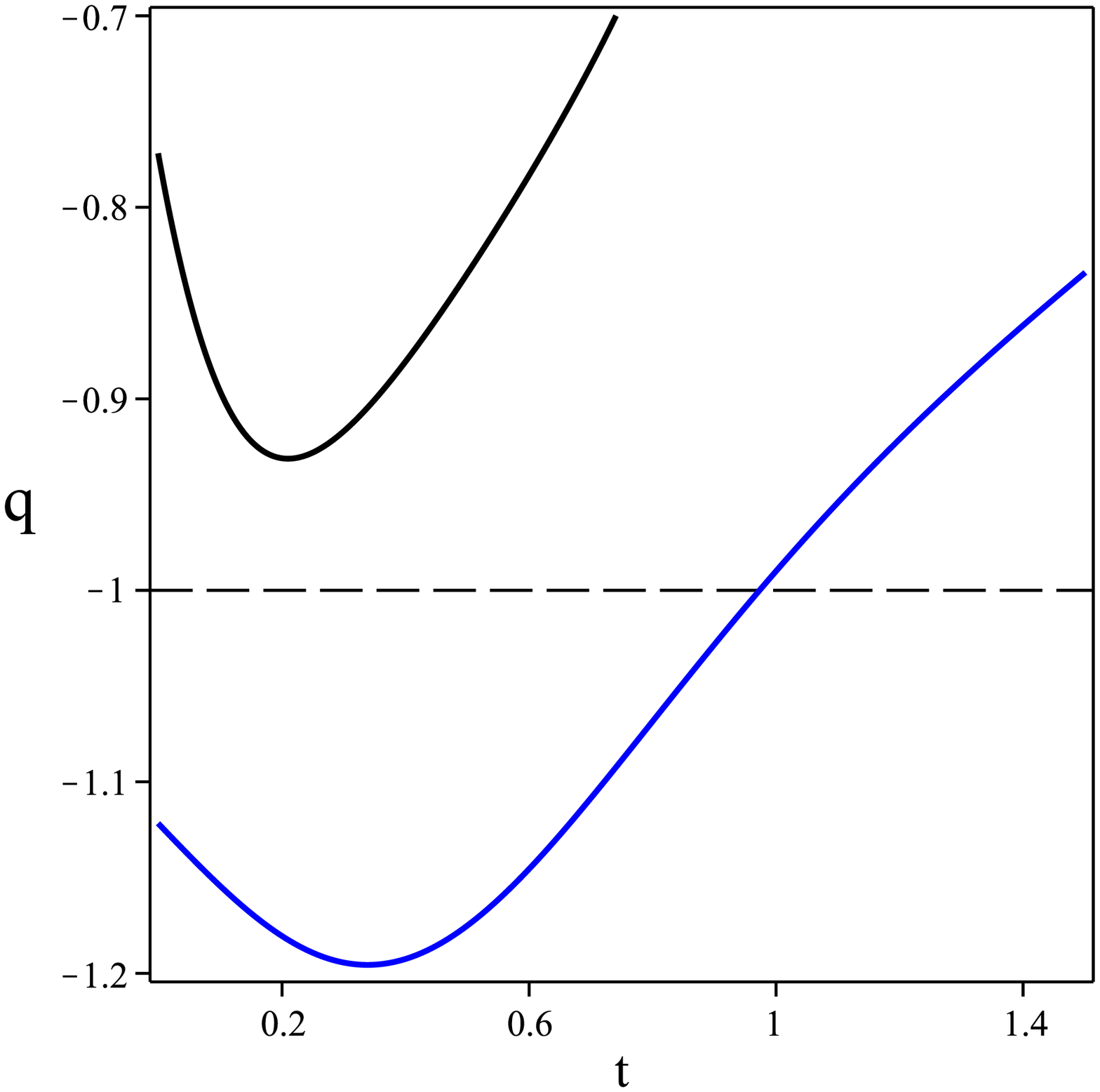}\hspace{0.1 cm}\\
Fig. 1:  The dynamical EoS parameter, $\omega_{eff}$, and deceleration parameter $q$ as a function of time.\\
The functions $f(\phi)=f_{0}\phi^\kappa$ and $V(\phi)=V_{0}\phi^\delta$, $V_0=20$.\\
For blue: $\kappa=1$, $\delta=-4$, $f_0=-0.05$. For black: $\kappa=1$, $\delta=-2$,  $f_0=0.1$ \\
I.Cs. $\phi(0)=-2$, $\dot{\phi}(0)=-1$, (blue)$a(0)=2$, $\dot{a}(0)=0.5$, (black)$a(0)=1$, $\dot{a}(0)=1.1$\\
\end{tabular*}\\

As one can see, in Fig. 1), for $\kappa=4$, $\delta=-2$, the phantom crossing occurs. Also, it shows that the behavior of the deceleration parameter is similar to $\omega_{eff}$, as expected. However, these graphs show only the dynamical behavior of these parameters with no observationally validated evidence. In the following, by utilizing stability analysis of the model we revisit the model and further explore the observability of it by fitting the model to the experimental data.

\section{perturbation and Stability Analysis}

In this section, we study the structure of the dynamical system via  phase plane analysis,
by introducing the following dimensionless variables,
\begin{eqnarray}\label{defin}
 \chi={\frac{\rho_{m}f}{3 H^{2}\phi}},\ \ \zeta={\frac{\dot{\phi}}{H\phi}} ,\ \ \eta^2=\frac{V}{3H^{2}\phi}.
\end{eqnarray}
Using equations (\ref{fried1})-(\ref{phiequation}), the evolution equations of these variables become,
\begin{eqnarray}
\chi '&=&\chi\{\frac{-3(1+\gamma)}{2}+[\frac{\kappa(1-3\gamma)-20}{8}]\zeta+\frac{\omega}{2}\zeta^{2}
+[\frac{3(1-3\gamma)(2-\delta)}{4(3+2\omega)}\label{kai}\\&+&
\frac{3(1+\gamma)}{2}]\chi^{2}+\frac{3\delta-6}{3+2\omega}\eta^{2})\},\ \nonumber\\
\zeta '&=&\zeta\{-3-3\zeta+\frac{\omega}{2}\zeta^{2}+[\frac{3(1-3\gamma)(2-\delta)}{4(3+2\omega)}+
\frac{3(1+\gamma)}{2}]\chi^{2}+\frac{3\delta-6}{3+2\omega}\eta^{2}\}\\&+&\frac{3(1-3\gamma)(2-\delta)}{2(3+2\omega)}\chi^{2}+\frac{6\iota-12}{3+2\omega}\eta^{2},\ \nonumber\\
\eta '&=&\eta\{\frac{\delta-5}{2}\zeta+\frac{\omega}{2}\zeta^{2}+[\frac{3(1-3\gamma)(2-\delta)}{2(3+2\omega)}+
\frac{3(1+\gamma)}{2}]\chi^{2}+\frac{3\delta-6}{3+2\omega}\eta^{2}\}, \label{zeta}
\end{eqnarray}
where prime in here and from now on is taken to be derivative with respect to $N = ln a$ , together with the Friedmann constraint equation (\ref{fried1}) which becomes
\begin{eqnarray}\label{constraint}
\chi^{2}-\zeta+\frac{\omega}{6}\zeta^{2}-\eta^{2}=1.
\end{eqnarray}
In term of the new dynamical variable, we also have,
\begin{eqnarray}
\frac{\dot{H}}{H^{2}}=2\zeta-\frac{\omega}{2}\zeta^{2}-[\frac{3(1-3\gamma)(2-\delta)}{2(3+2\omega)}-
\frac{3(1+\gamma)}{2}]\chi^{2}-\frac{3\delta-6}{3+2\omega}\eta^{2}.\label{hhdot}
\end{eqnarray}
Using the constraint (\ref{constraint}), the three equations (\ref{kai})-(\ref{zeta}) now reduce to the following two equations:
\begin{eqnarray}
\chi '&=&\chi\{\frac{-3(1+\gamma)}{2}+[\frac{\kappa(1-3\gamma)-20}{8}]\zeta+\frac{\omega}{2}\zeta^{2}
+[\frac{3(1-3\gamma)(2-\delta)}{4(3+2\omega)}\label{kai1}\\&+&
\frac{3(1+\gamma)}{2}]\chi^{2}+\frac{3\delta-6}{3+2\omega}(-1+\chi^{2}-\zeta+\frac{\omega}{6}\zeta^{2})\},\ \nonumber\\
\zeta '&=&\zeta\{-3-3\zeta+\frac{\omega}{2}\zeta^{2}+[\frac{3(1-3\gamma)(2-\delta)}{4(3+2\omega)}+
\frac{3(1+\gamma)}{2}]\chi^{2}\label{zeta1}\\&+&\frac{3\delta-6}{3+2\omega}(-1+\chi^{2}-\zeta+\frac{\omega}{6}\zeta^{2})\}
+\frac{3(1-3\gamma)(2-\delta)}{4(3+2\omega)}\chi^{2}+\frac{6\delta-12}{3+2\omega}(-1+\chi^{2}-\zeta+\frac{\omega}{6}\zeta^{2}).\ \nonumber
\end{eqnarray}
It is more convenient to investigate the properties of the dynamical system equations (\ref{kai1}) and (\ref{zeta1})
than equations (\ref{kai})-(\ref{zeta}). by calculating the critical points ( or fixed points) we study
the stability of these points. Critical points are always exact constant solutions in the
context of autonomous dynamical systems. These points are often the extreme points of
the orbits and therefore describe the asymptotic behavior of the system. In the following, we find fixed points by simultaneously solving $\chi '=0$ and $\zeta '=0$. Substituting
linear perturbations $\chi'\rightarrow \chi'+\delta \chi'$, $\zeta'\rightarrow \zeta'+\delta \zeta'$ about the critical points into the two independent equations (\ref{kai1}) and (\ref{zeta1}), to the first
orders in the perturbations, yields two eigenvalues $\lambda_{i} (i=1,2)$. Stability requires the real part of the eigenvalues to be negative. Solving the above equations we find seven fixed points which some of them explicitly depend on $\gamma$, $\kappa$ and $\delta$, as illustrated in tables I to III, for $\gamma=0$, $\gamma=1/3$ and $\gamma=-1$ respectively.\\

\begin{table}[ht]
\caption{critical points for $\gamma=0 $} 
\centering 
\begin{tabular}{c c c c c c c c} 
\hline\hline 
points  &  P1  & P2 \ & P3 \ & P4 \ & P5 \ & P6 \ & P7  \\ [4ex] 
\hline 
$\chi$ & $0 $ & 0 & 0 &$\frac{A}{\sqrt{3}(\kappa+8)}$ & -$\frac{A}{\sqrt{3}(\kappa+8)}$ & $ -\frac{B}{-\kappa+4\delta}$
& $ -\frac{B}{-\kappa+4\delta} $\\ 
\hline 
$\zeta $&$ 3+\sqrt{15}$& $3-\sqrt{15} $&$\frac{4-2\delta}{\delta+3}$ &$-\frac{2\kappa-4}{\kappa+8}$ & $-\frac{2\kappa-4}{\kappa+8} $ &$ -\frac{12}{-\kappa+4\delta}$ & $ -\frac{12}{-\kappa+4\delta}$  \\
\hline 
\end{tabular}
\label{table:1} 
\end{table}
where $A=\sqrt{-20\kappa-280+\kappa^{2}}$ and $B=\sqrt{-16\delta^{2}+56\delta+4\delta\kappa-8\kappa+72}$ \\
\begin{table}[ht]
\caption{critical points for $\gamma=\frac{1}{3} $} 
\centering 
\begin{tabular}{c c c c c c c c} 
\hline\hline 
points  &  P1  & P2 \ & P3 \ & P4 \ & P5 \ & P6 \ & P7  \\ [4ex] 
\hline 
$\chi$ & $0 $ & 0 & 0 &-1 & 1  & $ \frac{\sqrt{\delta^{2}-4\delta-6}}{\delta}$
& $ -\frac{\sqrt{\delta^{2}-4\delta-6}}{\delta} $\\ 
\hline 
$\zeta $&$ 3+\sqrt{15}$& $3-\sqrt{15} $&$\frac{4-2\delta}{\delta+3}$ & 0 & 0  &$ -\frac{4}{\delta}$ & $ -\frac{4}{\delta}$ \\
\hline 
\end{tabular}
\label{table:1} 
\end{table}
\begin{table}[ht]
\caption{critical points for $\gamma=-1 $} 
\centering 
\begin{tabular}{c c c c c c c c} 
\hline\hline 
points  &  P1  & P2 \ & P3 \ & P4 \ & P5 \ & P6 \ & P7  \\ [4ex] 
\hline 
$\chi$& $0 $ & 0 & 0 &$ \sqrt{\frac{\delta-2}{\delta-\kappa}}$&-$ \sqrt{\frac{\delta-2}{\delta-\kappa}}$ & -$\frac{C}{3\kappa+9}$
& $\frac{C}{3\kappa+9}$\\ 
\hline 
$\zeta$ &$ 3+\sqrt{15}$& $3-\sqrt{15} $&$\frac{4-2\delta}{\delta+3}$ & 0 & 0  &$ \frac{4-2\kappa}{\kappa+3}$ & $  \frac{4-2\kappa}{\kappa+3}$ \\
\hline 
\end{tabular}
\label{table:2} 
\end{table}

 where $C=\sqrt{165- 60\kappa -15\kappa^{2}}$.\\

In the following we will investigate the stability of the model
with respect to the above three specific choices of the EoS parameter for the matter in the universe, i.e. $\gamma=0$, $\gamma=1/3$ and $-1$.

\subsection{Stability for $\gamma=0$}
In the case of $\gamma=0$, the stability properties of the seven critical points in this system are shown in the following:
\begin{eqnarray}
P1&:&\ \lambda_{1}=(3+\sqrt{15})(-2+\sqrt{15}+\delta),\ \lambda_{2}=\frac{1}{8}(3+\sqrt{15})(-5+\sqrt{15}+\kappa),\ \mbox{stable for} \nonumber \\&& \ \ \ \delta<2-\sqrt{15} , \kappa<5-\sqrt{15}\nonumber\\
P2&:&\ \lambda_{1}=(3-\sqrt{15})(-2-\sqrt{15}+\delta),\ \lambda_{2}=\frac{1}{8}(3-\sqrt{15})(-5-\sqrt{15}+\kappa),\ \mbox{stable for} \nonumber \\&& \ \ \ \delta>2+\sqrt{15} , \kappa>5+\sqrt{15}\nonumber\\
P3&:&\ \lambda_{1}=(-11-4\delta+\delta^2)/(3+\delta),\ \lambda_{2}=(4\delta^2-14 \delta-\delta\kappa+2\kappa-18)/(12+4\delta) , \mbox{stable for} \nonumber\\
&&\ i )\delta<-3,\varepsilon < \kappa\ ,
\ \ ii )2<\delta<2+\sqrt{15} ,\varepsilon < \kappa\
  iii )2-\sqrt{15}<\delta<2 ,\kappa<\varepsilon\ \ iiii)\forall \kappa, \delta=2\nonumber\\
P4,P5&:&\lambda_{1}=(\kappa^2-4\kappa-56)/(4\kappa+32), \lambda_{2}=(\kappa^2+4\kappa-4\delta\kappa+8\delta+48)/(2\kappa+16),\ \ \mbox{stable for }\nonumber\\
&&\ i )\kappa<-8,\mu< \delta\ ,
\ \ ii )2<\kappa<2+\sqrt{60} ,\mu< \delta\  iii )2-\sqrt{60}<\kappa<2 ,\delta<\mu\nonumber\\
P6,P7 &:& \ \lambda_{1,2}=\frac{-30\delta+30+15\kappa\pm\sqrt{Q}}{-10\kappa+40\delta}\ \mbox{stable for some values of $\delta$ and $\kappa$, }
\end{eqnarray}
where $\varepsilon=2(2\delta^2-7\delta-9)/(\delta-2),\ \ \     \mu=(\kappa^2+4\kappa+48)/(4\kappa-8)$, and\\
$Q=-1500\delta^2+22680\delta-420\kappa\delta+26820+180\kappa+525\kappa^2+780\delta\kappa^2+
30\kappa^3\delta-60\kappa^3+480\delta^3\kappa-960\delta^3-1920\kappa\delta^2-240\delta^2\kappa^2.$

We see that all the critical points are stable for the given conditions on the stability parameters $\kappa$ and $\delta$.

In Fig. 2), the attraction of trajectories to the critical points P1 to P7 in the phase plane is shown for the given conditions on $\kappa$ and $\delta$. Fig. 2a), belongs to the critical point P1. Fig.2b) is an example of the stability of the critical point P3 that satisfies the condition i). In Fig 2c) since there are two critical points P4 and P5 with the same stability conditions, in here, condition i) gives us two stable critical points. The same argument is applies to the Fig 2d) with respect to the points P6 and P7 where for some numerical values of $\delta$ and $\kappa$ the two eigenvalues become negative and critical points become stables. In addition, since for the critical point P1 and P3, $\chi=0$ and $\zeta\neq 0 $ for the stability conditions on $\delta$ and $\kappa$, these points corresponds to a scalar field dominated era.

\begin{tabular*}{2.5 cm}{cc}
\includegraphics[scale=.35]{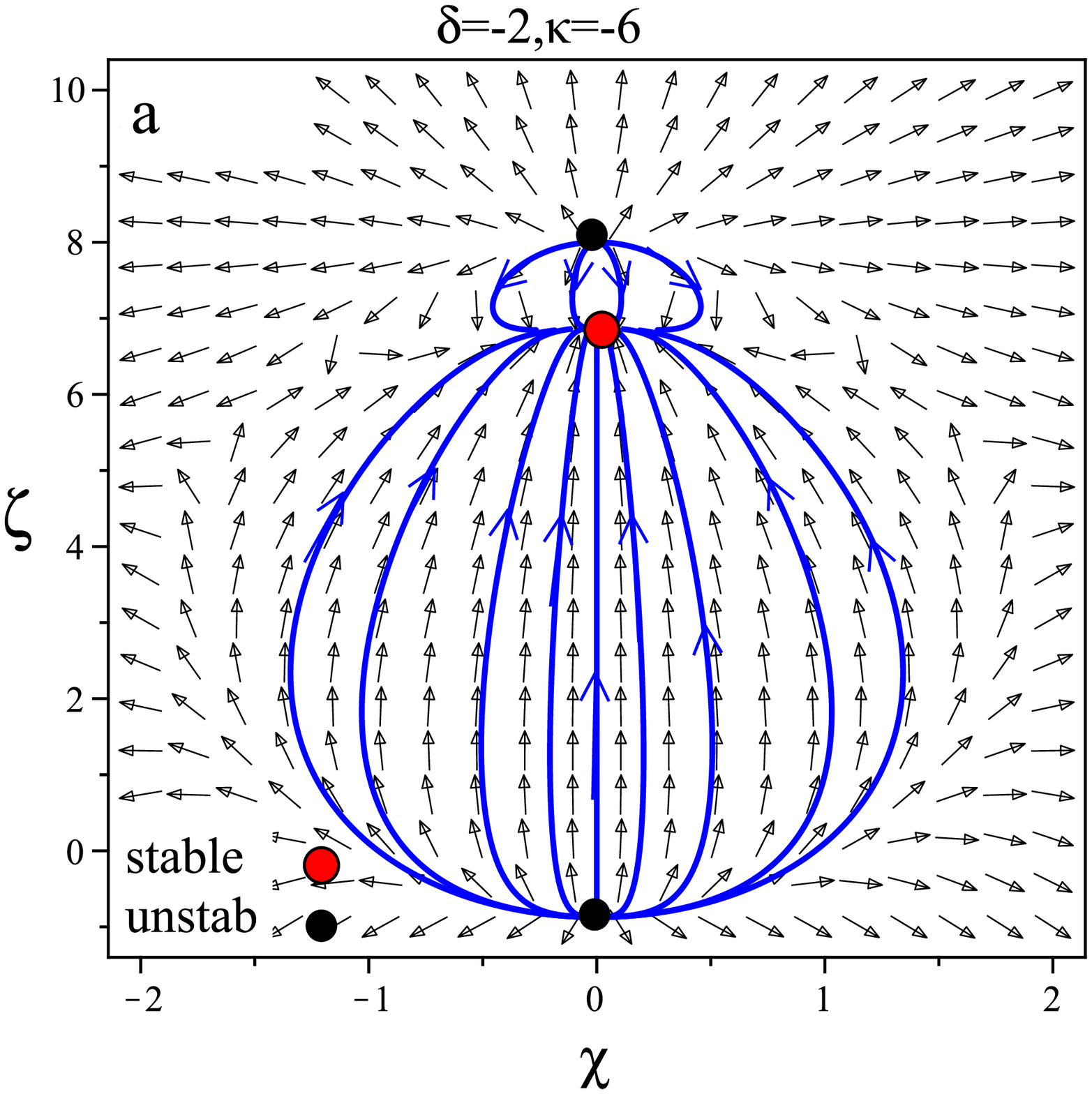}\hspace{0.1 cm}\includegraphics[scale=.35]{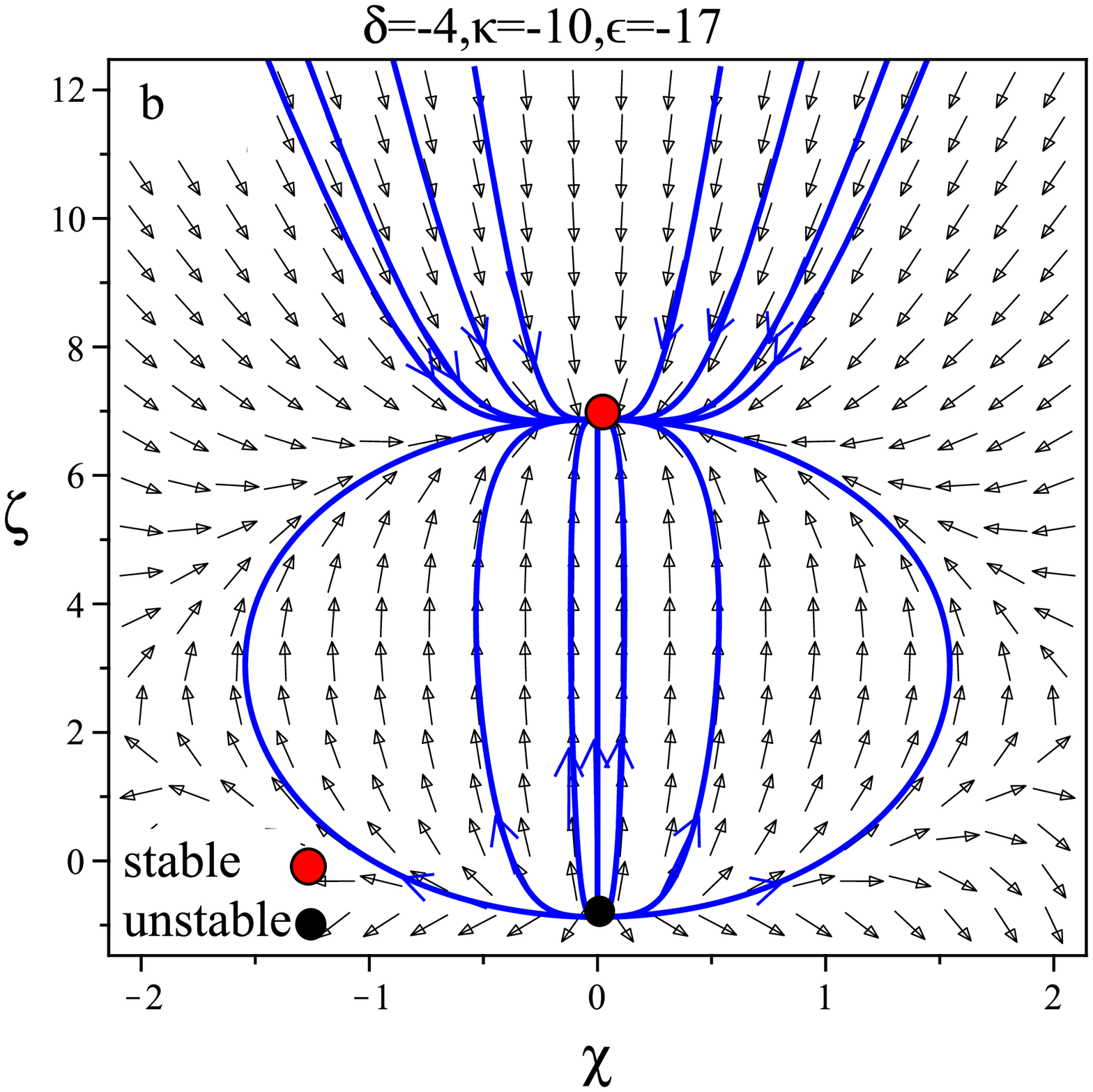}\hspace{0.1 cm}\\
\includegraphics[scale=.35]{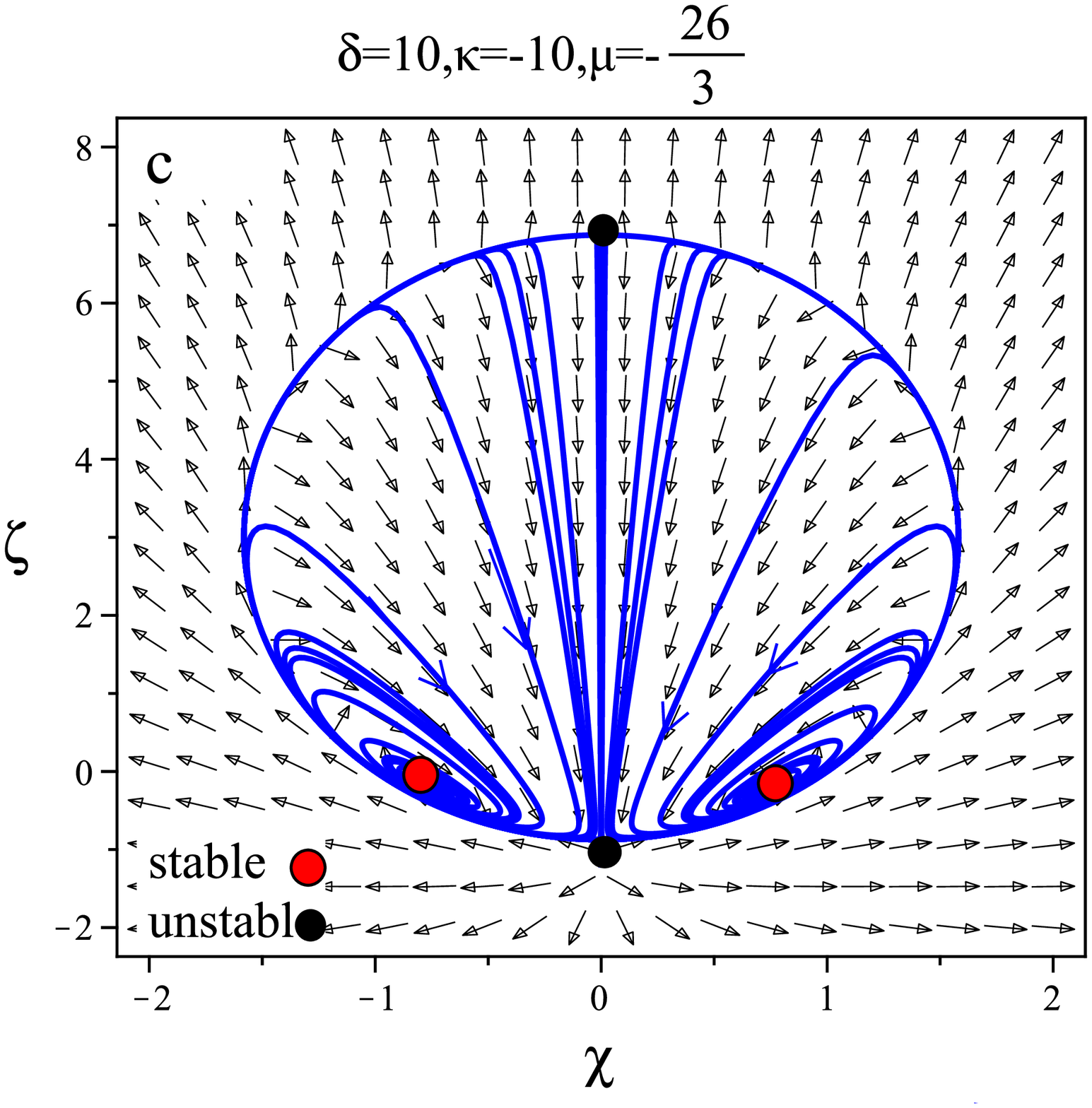}\hspace{0.1 cm}\includegraphics[scale=.35]{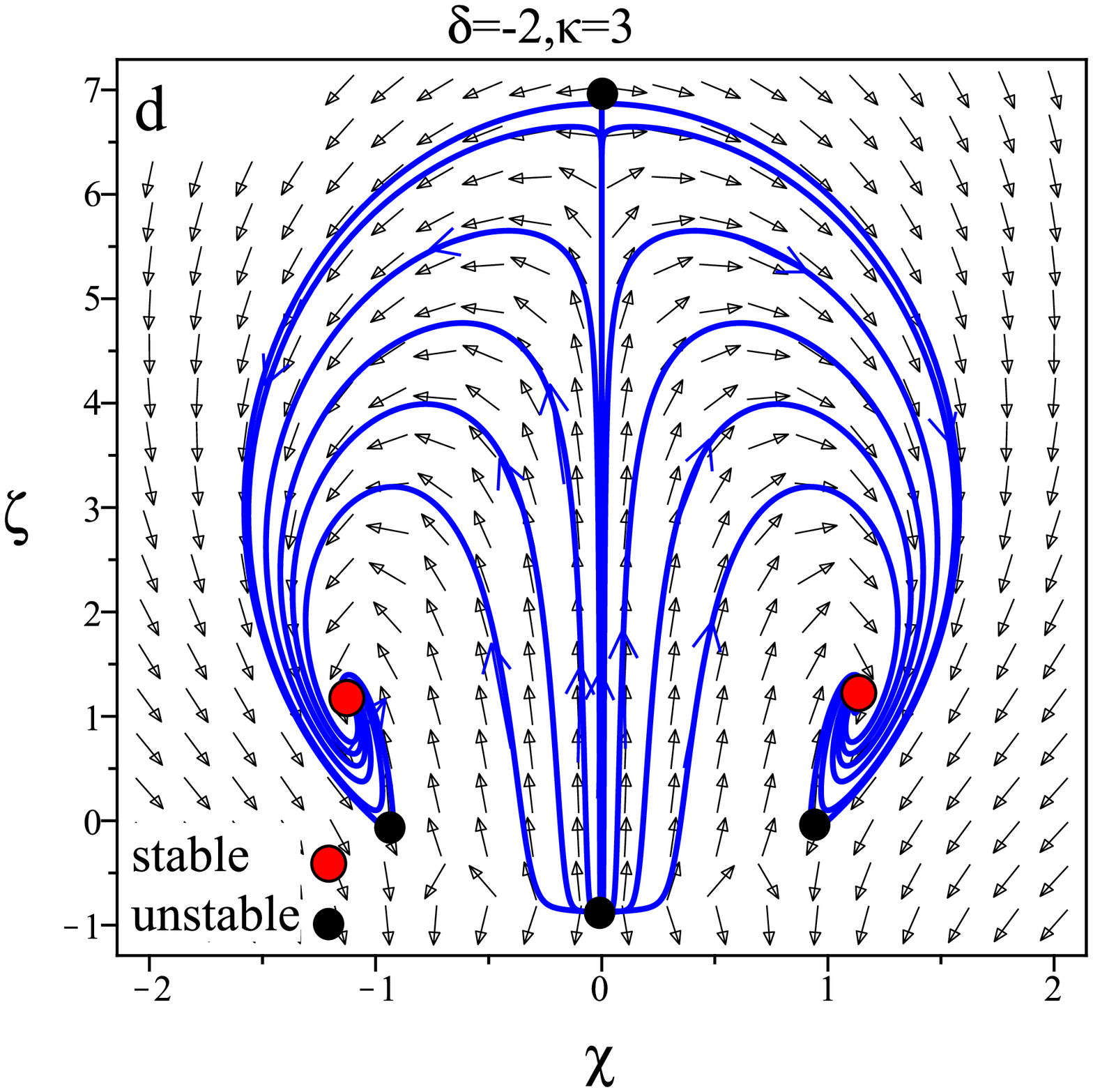}\hspace{0.1 cm}\\
Fig. 2:  The attractor property of the dynamical system in
the phase plane for the case $\gamma=0$.\\ The plots a) to d) are respectively examples of the critical points $P1$ to  $P7$.\\
\end{tabular*}\\

\subsection{Stability for $\gamma=\frac{1}{3}$}

In the case of $\gamma=\frac{1}{3}$, the stability properties of the seven critical points in this system are shown in the following:
\begin{eqnarray}
P1,P2&:&\ \lambda_{1}=(3\pm\sqrt{15})(2\pm\sqrt{15}+\delta),\ \lambda_{2}=\frac{5}{2}\pm\frac{\sqrt{15}}{2} ,\ \mbox{unstable}\ \nonumber\\
P3&:&\ \lambda_{1}=(-11-4\delta+\delta^2)/(3+\delta),\ \lambda_{2}=(\delta^2-4 \delta-6)/(3+\delta) , \mbox{stable for:}\nonumber\\
& &  \ i )\forall \kappa, \delta<-3\ ,  \ \ ii )\forall \kappa, 2-\sqrt{10}<\delta<2+\sqrt{10} \\
P4,P5&:&\lambda_{1}=4, \lambda_{2}=-1,\ \ \mbox{unstable (saddle point)}\nonumber\\
P6,P7 &:& \ \lambda_{1,2}=\frac{2-\delta\pm\sqrt{100+60\delta-15\delta^2}}{2\delta},\  \mbox{stable for} \ i)\forall \kappa, \delta>2+\sqrt{10}, \ ii)\forall \kappa, \delta<2-\sqrt{10} \nonumber
\end{eqnarray}
We explicitly find that the critical points $P3, P6, P7$ are stable for the given conditions on the stability parameters $\kappa$ and $\delta$.
In Fig. 3), the attraction of trajectories to the critical points in the phase plane is shown for the given conditions on $\kappa$ and $\delta$. In Fig. 3a), although the given stability parameter $\delta=4$ satisfies the condition i) of the critical point P3, since it also satisfies the conditions ii) of the critical points  P6 and P7, as can be seen, there are three stable points displayed in the graph. In addition, since for the critical point P3, $\chi=0$ and $\zeta\neq 0 $ for all $\delta$ and $\kappa$, this point corresponds to a scalar field dominated era.

\begin{tabular*}{2.5 cm}{cc}
\includegraphics[scale=.35]{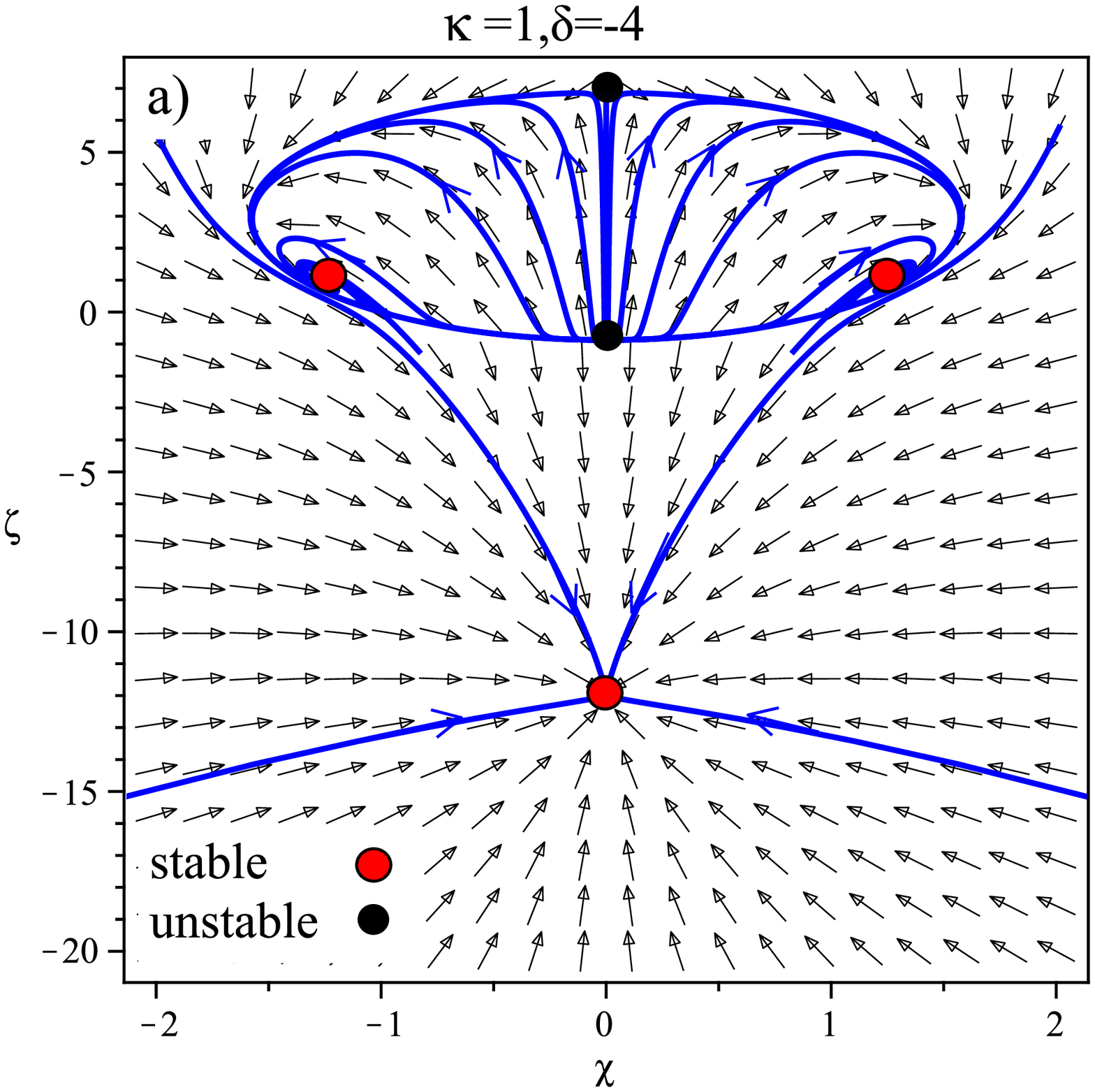}\hspace{0.1 cm}\includegraphics[scale=.35]{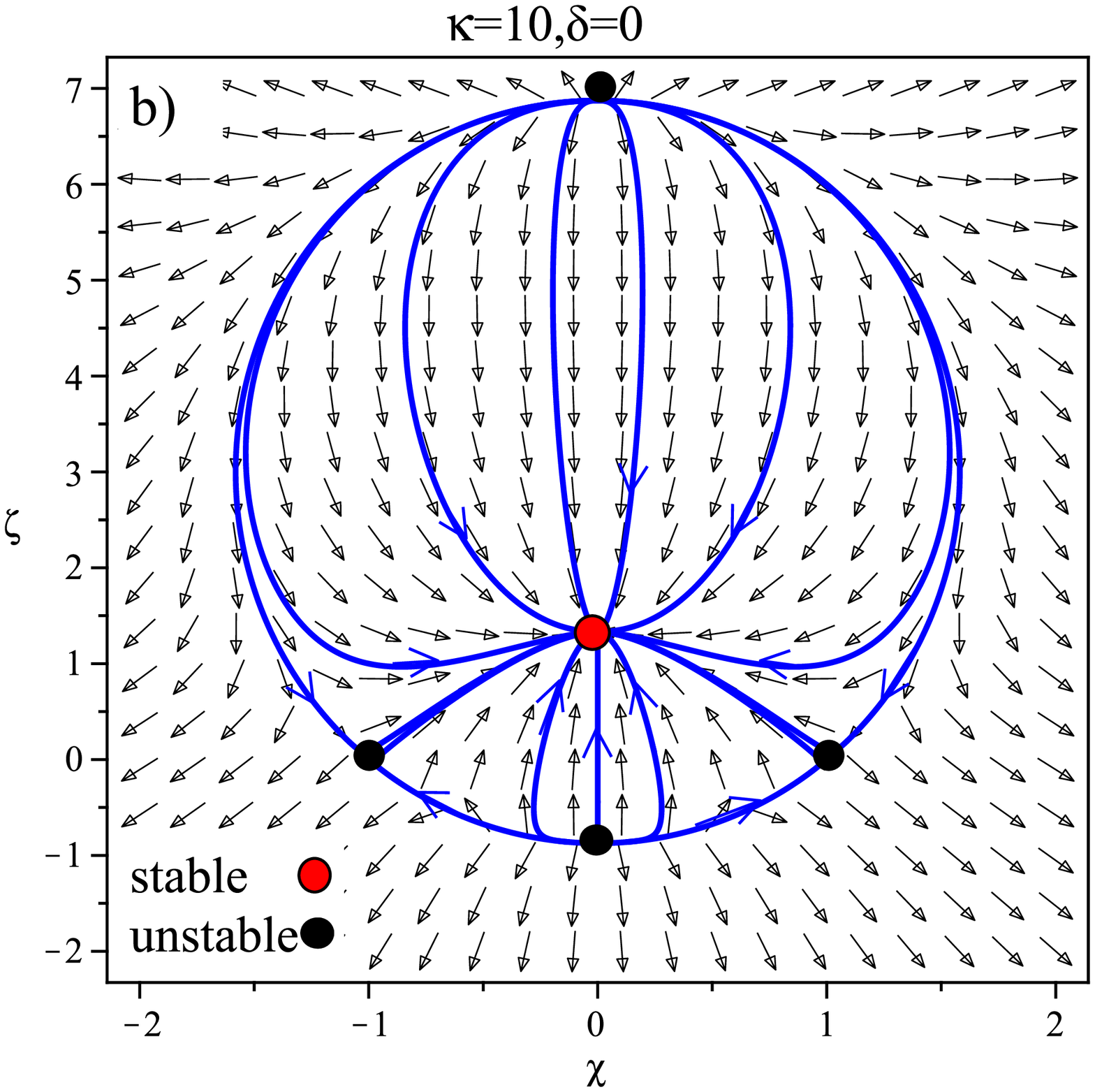}\hspace{0.1 cm}\\
\includegraphics[scale=.35]{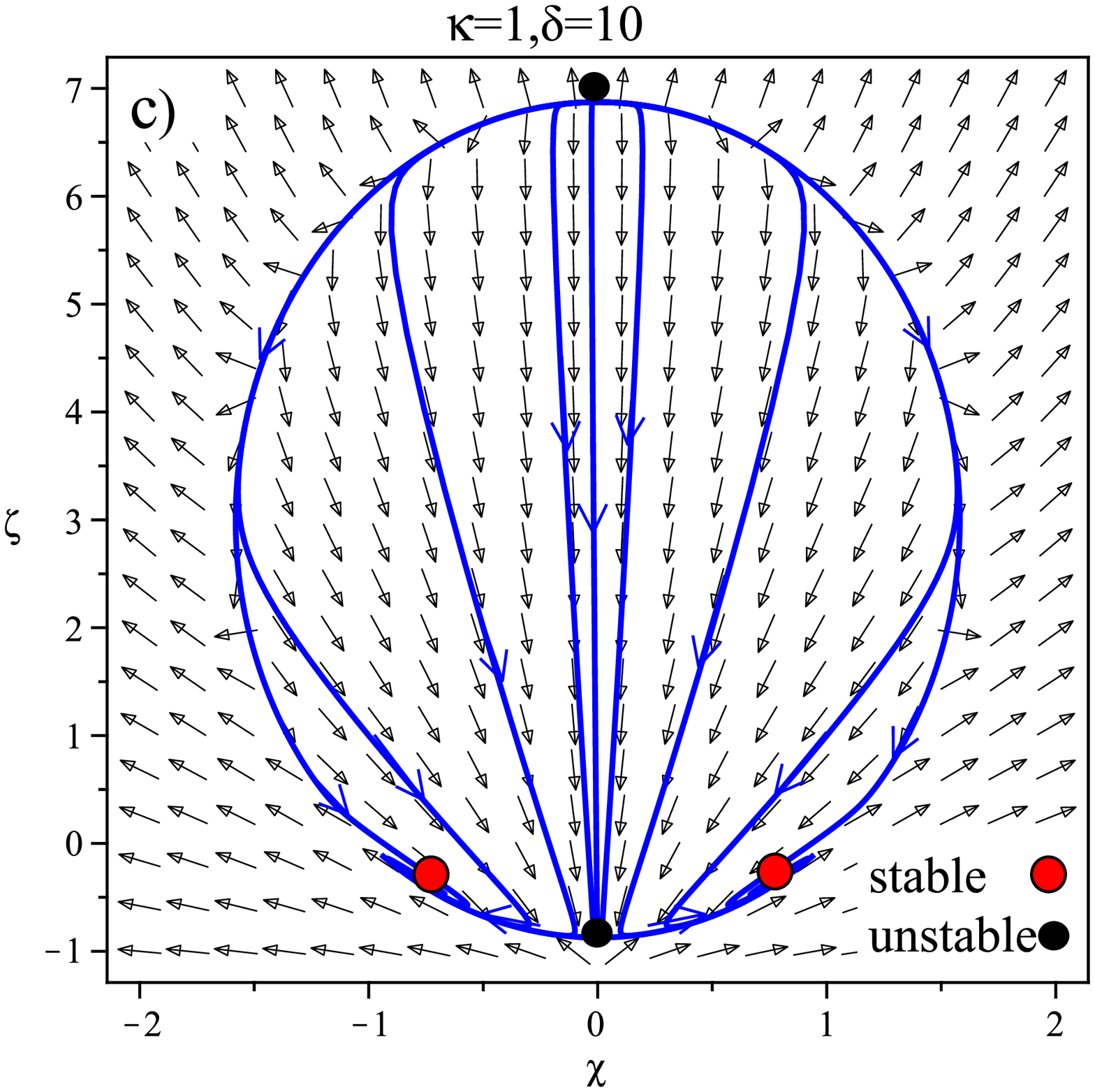} \hspace{0.1 cm}\includegraphics[scale=.35]{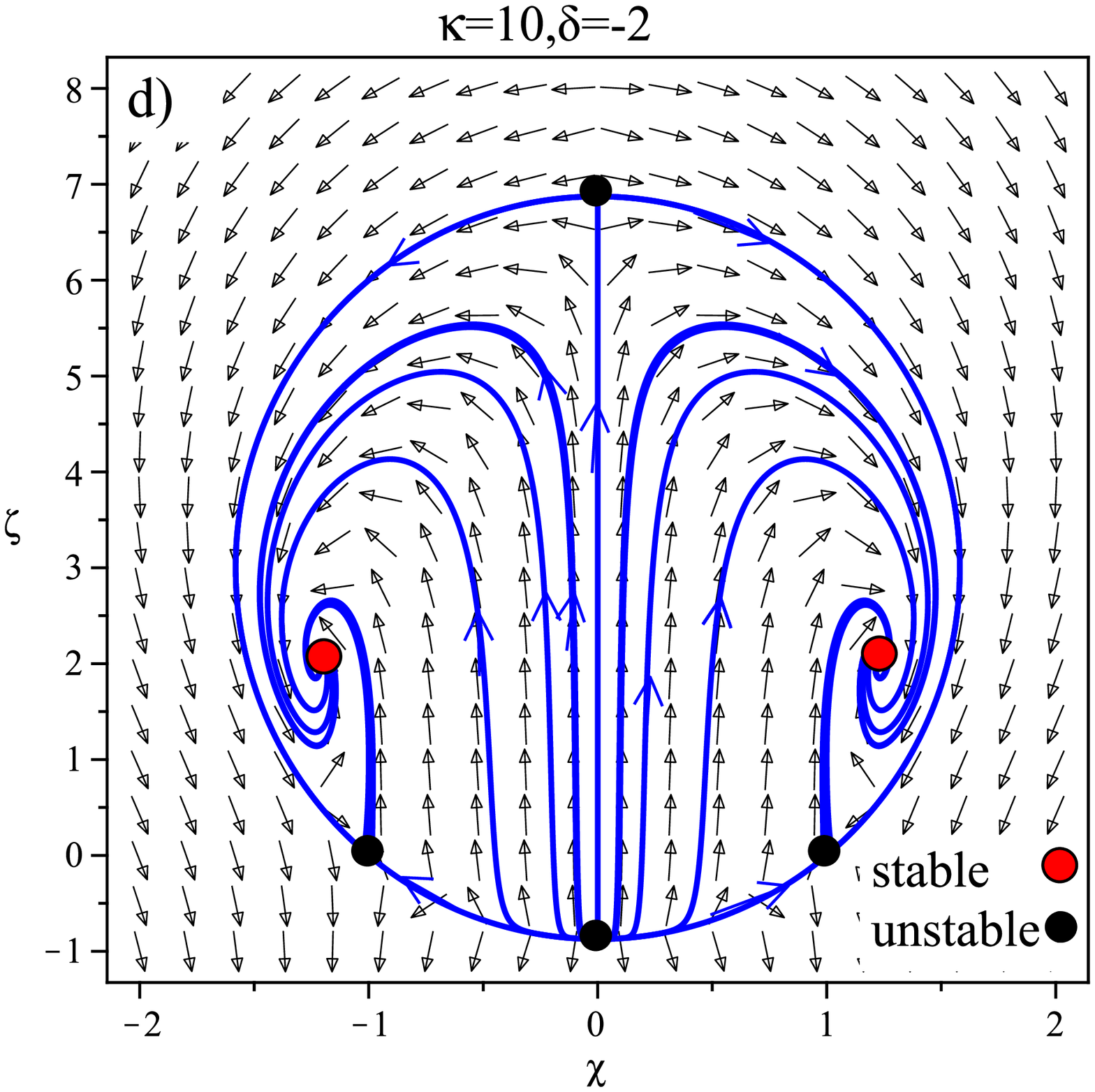}\hspace{0.1 cm}\\
Fig. 3:  The attractor property of the dynamical system in
the phase plane for the case $\gamma=1/3$.\\ The plots a) to d) are respectively examples of the critical points $P3, P6$ and $ P7$.\\
\end{tabular*}\\

\subsection{Stability for $\gamma=-1$}

In this case, we find the stability conditions for the seven fixed points in the system as shown in the following:
\begin{eqnarray}
P1 &:&\ \lambda_{1}=(3+\sqrt{15})(2+\sqrt{15}+\delta),\ \lambda_{2}=(3+\sqrt{15})(2+\sqrt{15}+\kappa),\ \mbox{stable for} \nonumber \\&&   \ \ \ \delta<2-\sqrt{15} , \kappa<2-\sqrt{15} \nonumber\\
P2&:&\ \lambda_{1}=(3-\sqrt{15})(2-\sqrt{15}+\delta),\ \lambda_{2}=(3-\sqrt{15})(2-\sqrt{15}+\kappa),\ \mbox{stable for} \nonumber \\&&   \ \ \ \delta<2-\sqrt{15} , \kappa<2-\sqrt{15} \nonumber\\
P3&:&\ \lambda_{1}=\frac{(-11-4 \delta+\delta^2)}{(3+\delta)},\ \lambda_{2}=\frac{(-2\delta+2\kappa+\delta^2-\kappa\delta)}{(3+\delta)},\  \mbox{stable for}\nonumber \\
&&i)  \delta<\kappa,\delta<-3 \ \ ii) \delta<\kappa,2<\delta<2+\sqrt{15}\ \ iii) \delta>\kappa,2-\sqrt{15}<\delta<2  \nonumber\\
P4,P5&:& \ \lambda_{1,2}= -\frac{3}{2}\pm\frac{1}{10}\sqrt{705-240\kappa+120\delta\kappa-240\delta}, \mbox{stable for}\ \ i) \delta<0,\kappa>\alpha \ \ ii) \delta>0,\kappa<\alpha\nonumber \\
\nonumber \  \ && iii) \delta<2,\kappa>\beta \ \ iiii) \delta>2,\kappa<\beta \nonumber\\
P6,P7&:& \ \lambda_{1}= \frac{-4 \kappa+\kappa^2-11}{\kappa+3},\lambda_{2}=\frac{-2(-\kappa^2+2\kappa+\kappa\delta-2\delta)}{\kappa+3}, \
\mbox{stable for} \nonumber \\ && \ \ i)\kappa<\delta,2<\kappa<2+\sqrt{15} \ \ ii)\kappa>\delta,2-\sqrt{15}<\kappa<2 \nonumber
\end{eqnarray}
where $alpha=\frac{2(-1+\delta)}{\delta},\ \ \beta=\frac{1}{8}\frac{(-47+16\delta)}{\delta-2}$

Again, we explicitly find that all of the critical points are stable by the given conditions on $\kappa$ and $\delta$.
In Fig. 4), the attraction of trajectories to the critical points P1 and P3 in the phase plane is shown for particular choice of $\kappa$ and $\delta$.
In Fig. 4b), similar to the argument we had in case $\gamma=1/3$, although the given stability parameters $\delta=-4$, $\kappa=10$ satisfies the condition i) of critical point P3, since it also satisfies the conditions iii) of the critical points  P4 and P5, there are three stable points displayed in the graph.  Also, Fig. 5) shows the attraction of trajectories to the critical points P4 to P7 in the phase plane for the conditions on $\kappa$ and $\delta$. Again, the extra stable points shown in Fig. 5 a), c), d), and f) correspond to the points satisfied by the stability conditions for other critical points. Moreover, since for the critical points P1, P2 and P3, $\chi=0$ and $\zeta\neq 0 $ for all $\delta$ and $\kappa$, these points correspond to a scalar field dominated era. Similarly, for the critical points P4 and P5, $\zeta =0 $ and $\chi \neq0$  for all $\delta$ and $\kappa$, therefore, these points correspond to either a matter dominated era or $f(\phi)$ dominated era.\\

\begin{tabular*}{2.5 cm}{cc}
\includegraphics[scale=.35]{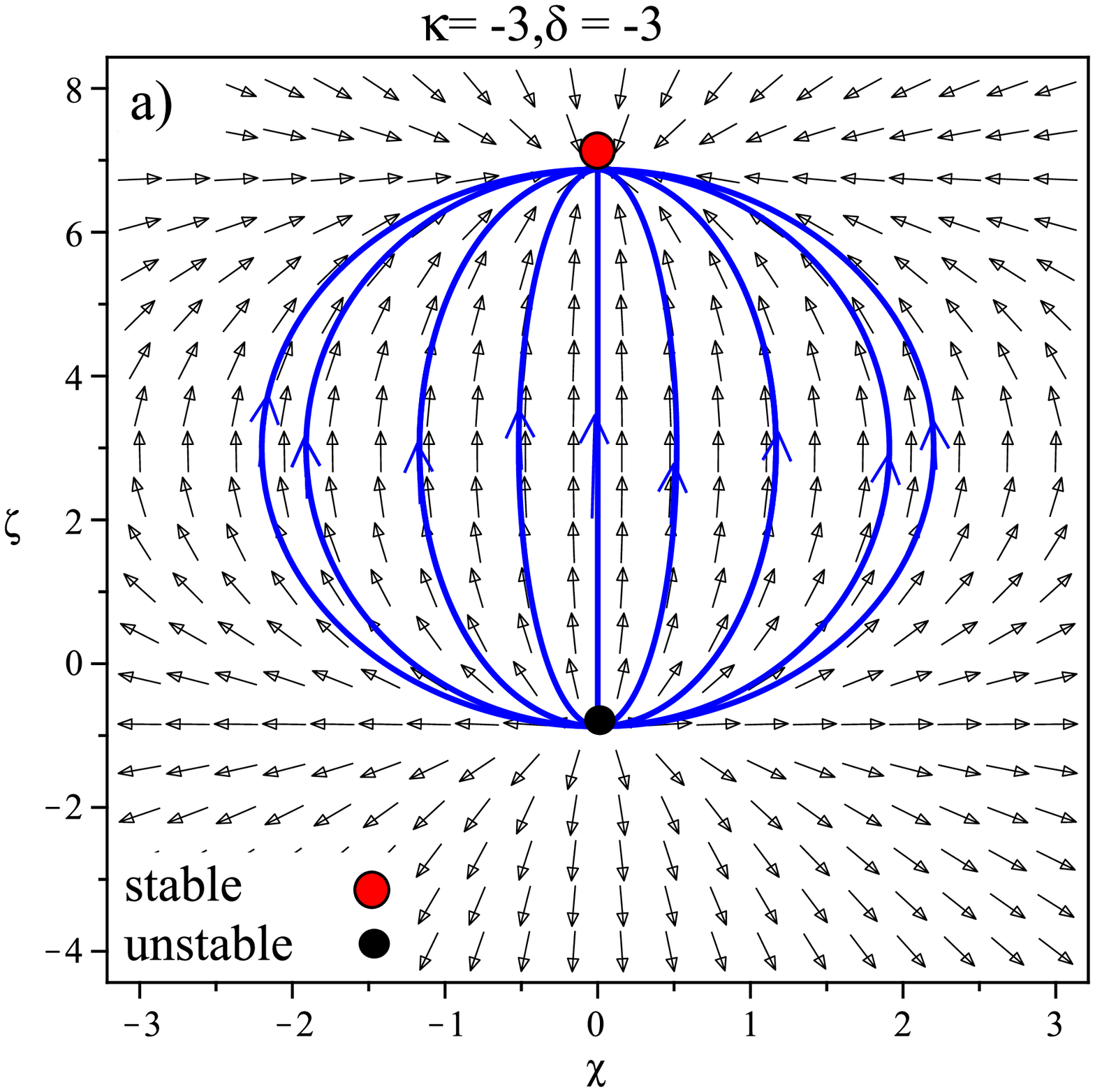}\hspace{0.1 cm}\includegraphics[scale=.35]{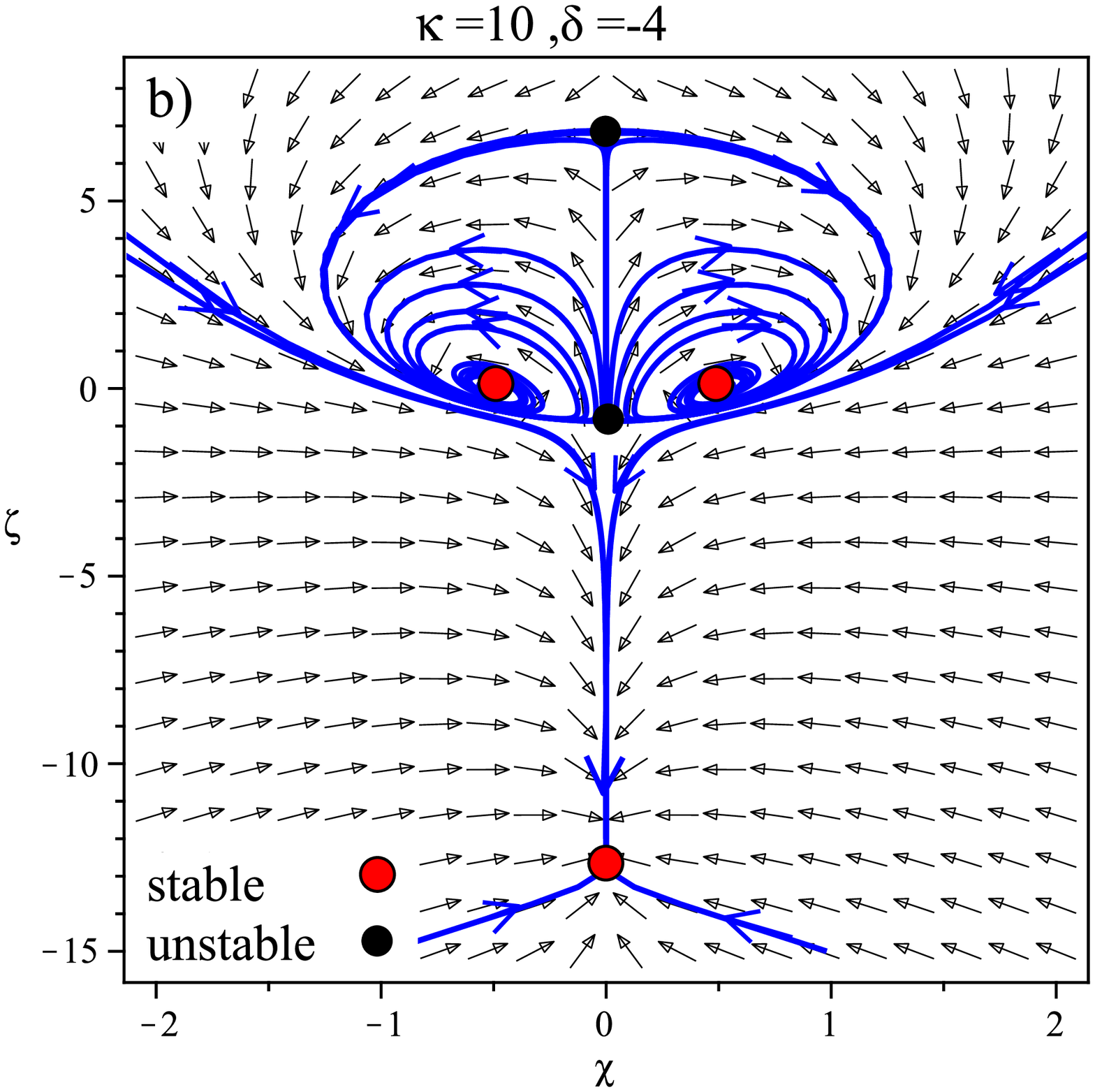}\hspace{0.1 cm}\\
\includegraphics[scale=.35]{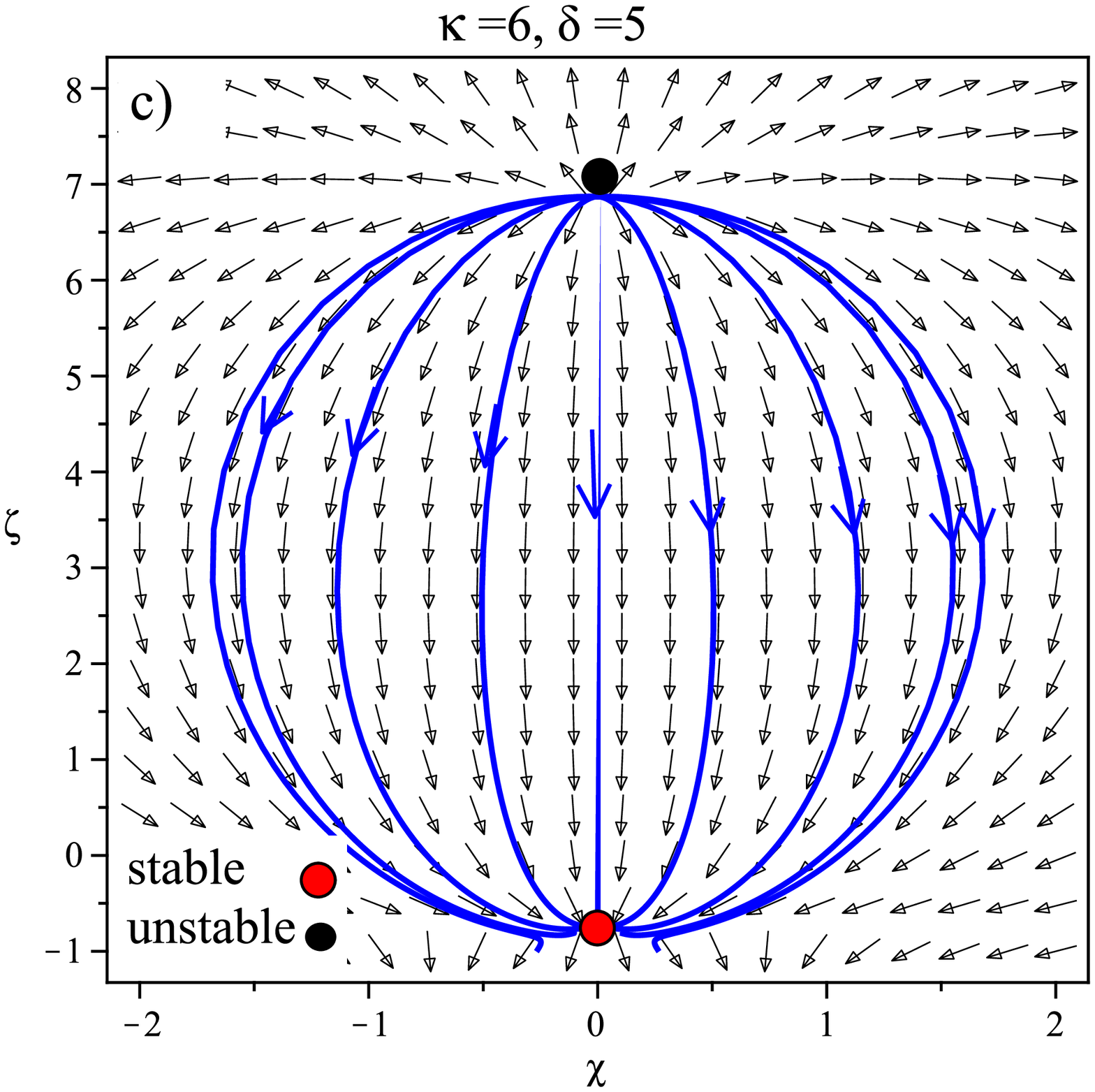}\hspace{0.1 cm}\includegraphics[scale=.35]{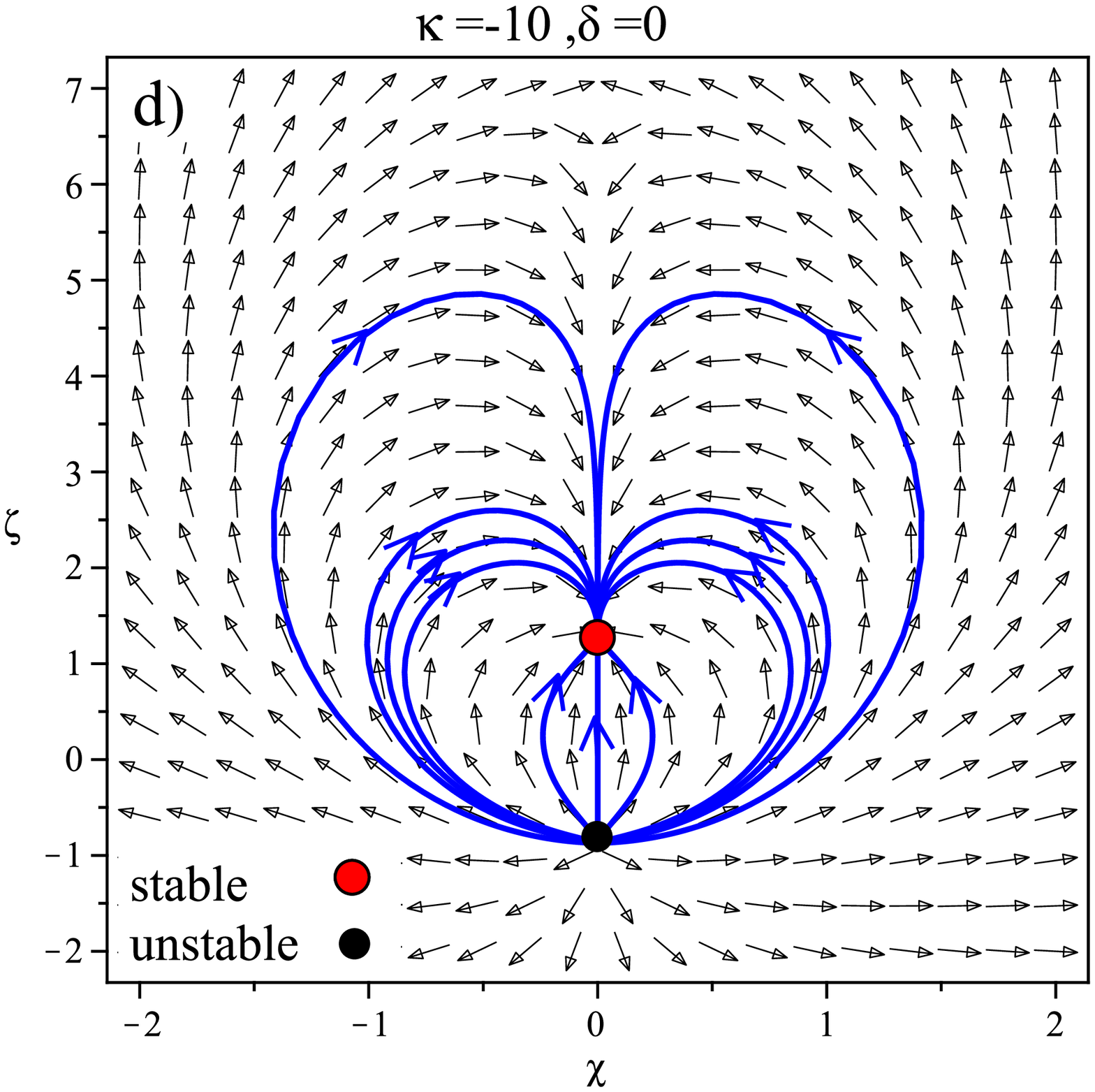}\hspace{0.1 cm}\\
Fig. 4:  The attractor property of the dynamical system in
the phase plane for the case $\gamma=-1$.\\ The plots a) to d) are respectively examples of the critical points $P1, P2$ and $P3$.\\
\end{tabular*}\\

\begin{tabular*}{2.5 cm}{cc}
\includegraphics[scale=.35]{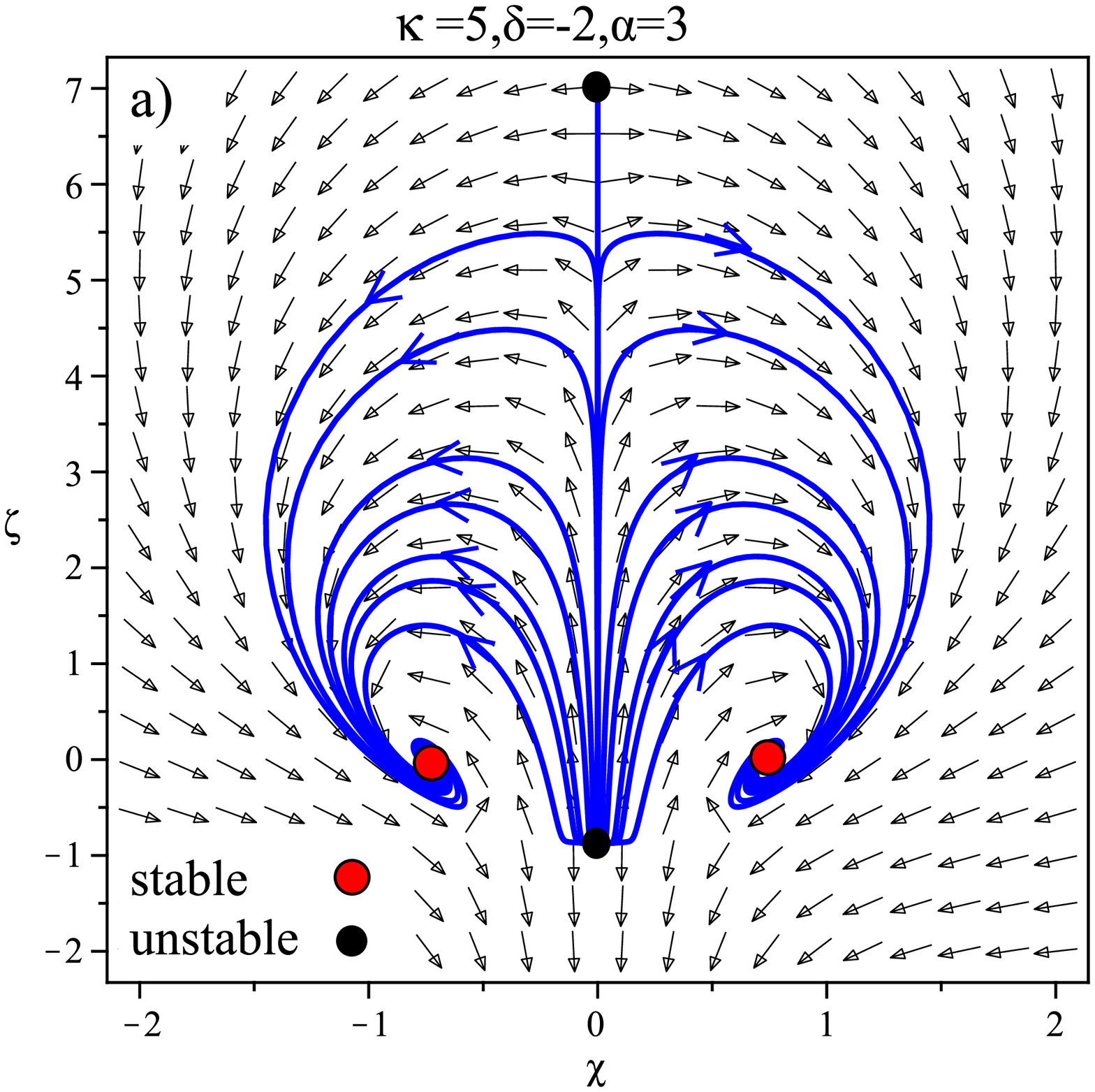}\hspace{0.1 cm}\includegraphics[scale=.35]{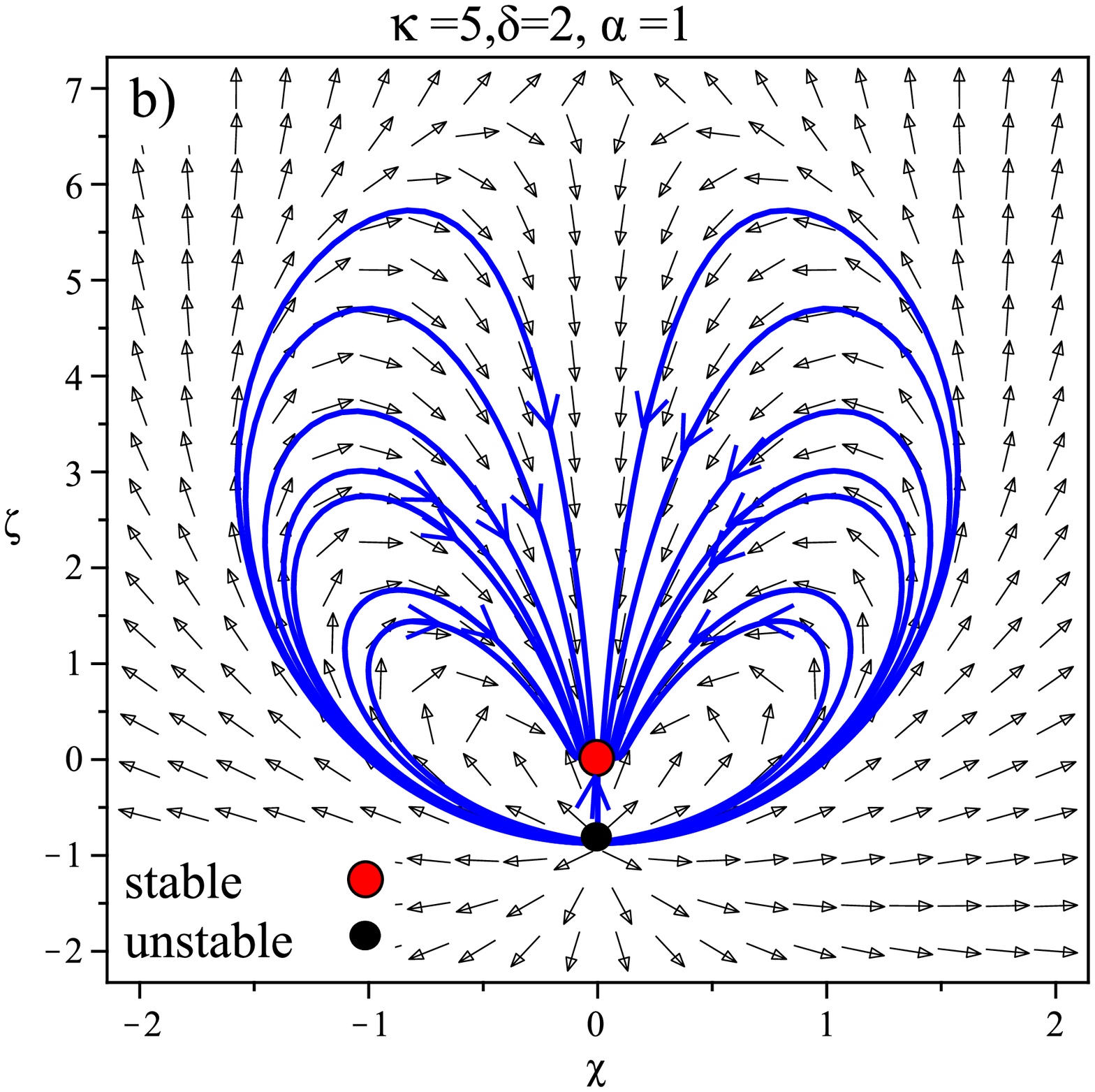}\hspace{0.1 cm}\\
\includegraphics[scale=.35]{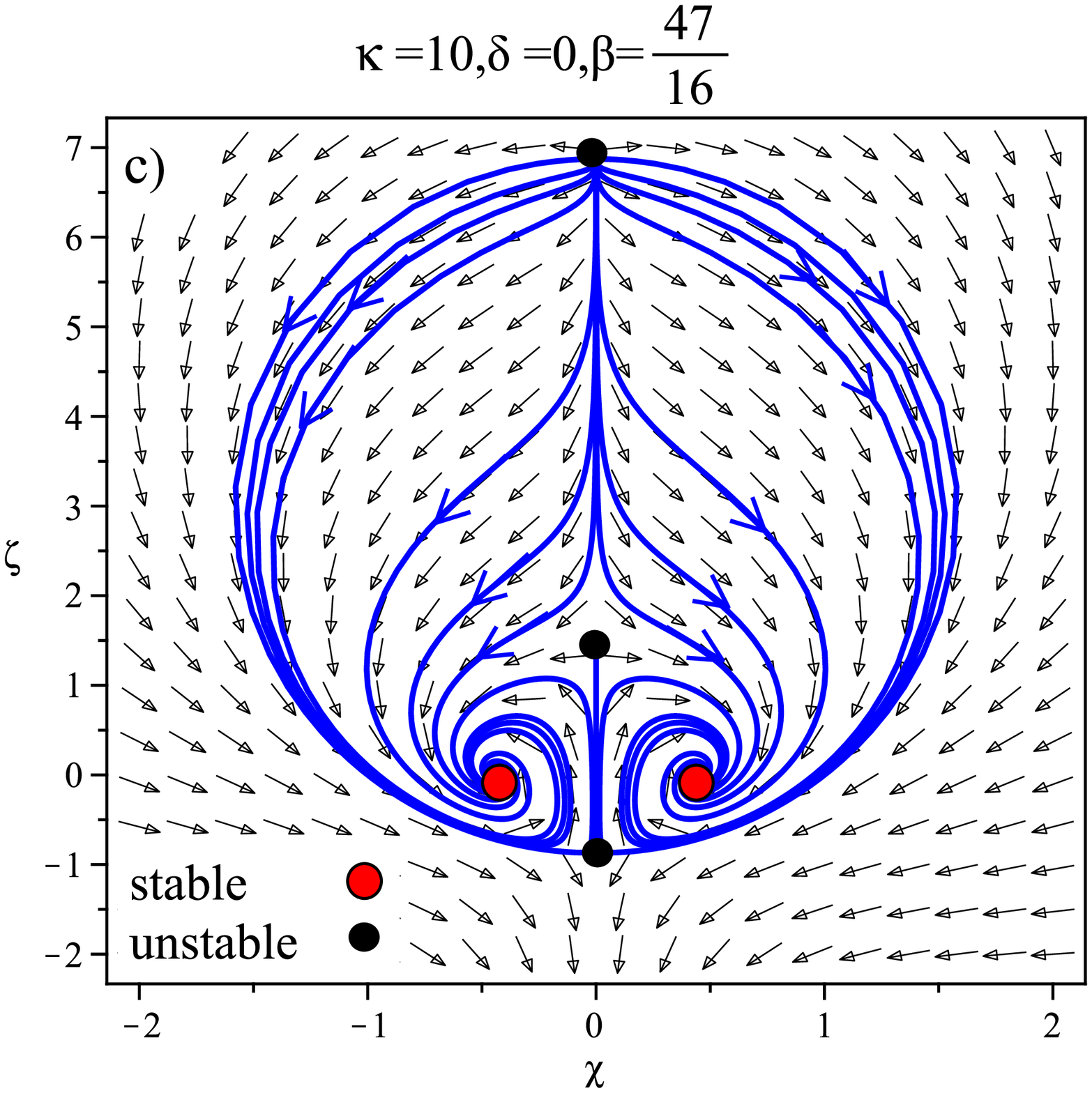}\hspace{0.1 cm}\includegraphics[scale=.35]{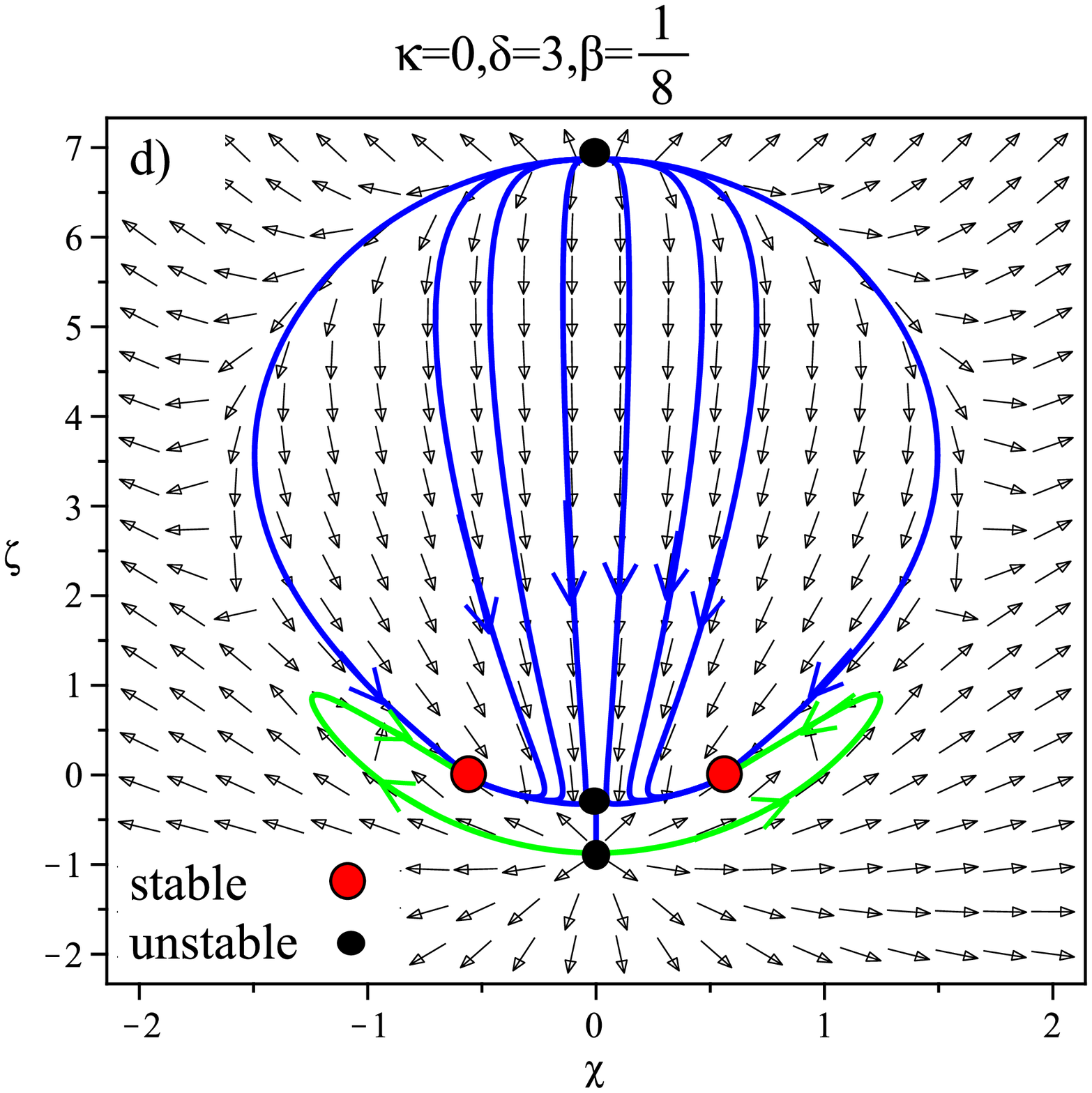}\hspace{0.1 cm}\\
\includegraphics[scale=.38]{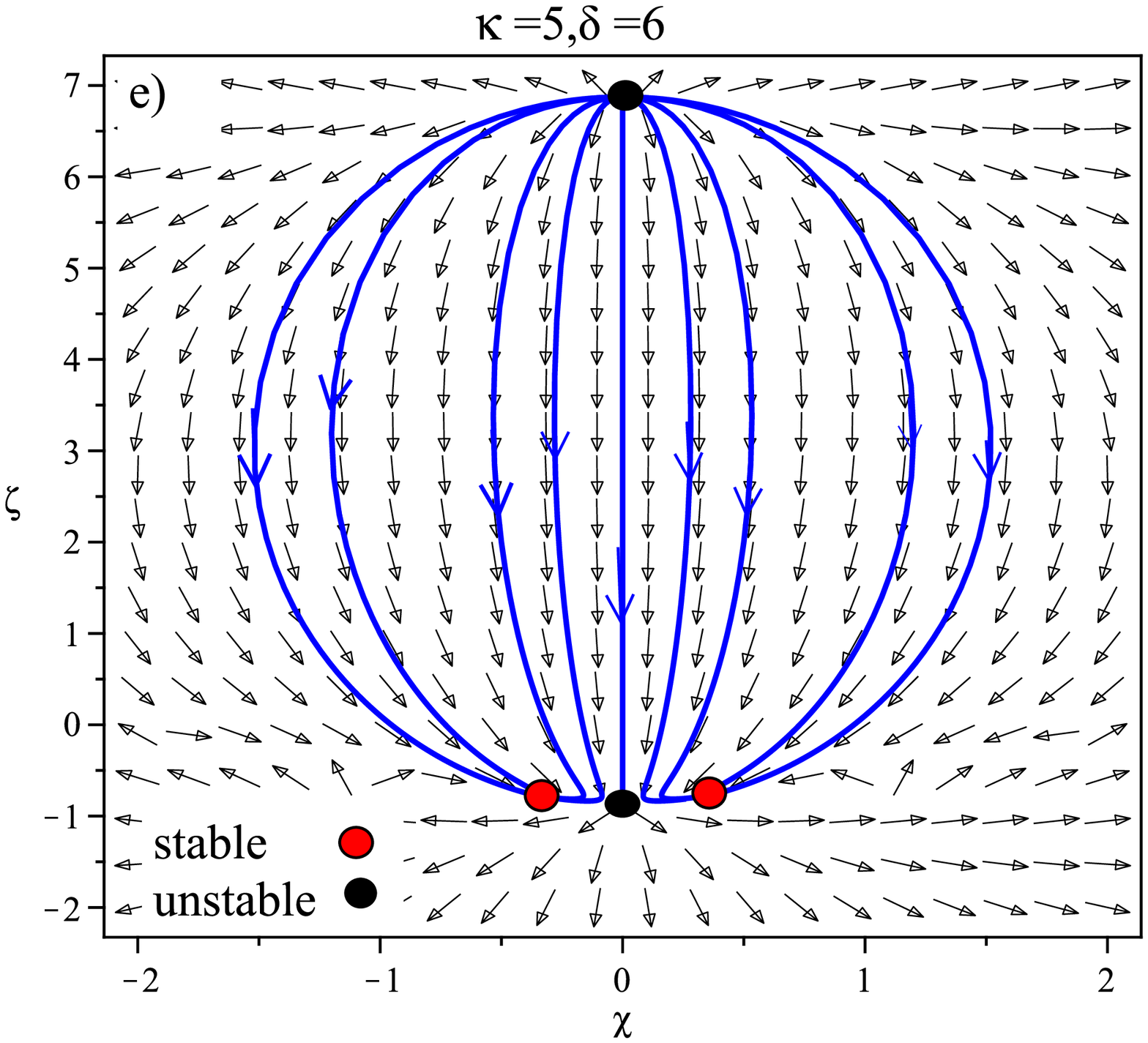}\hspace{0.1 cm}\includegraphics[scale=.35]{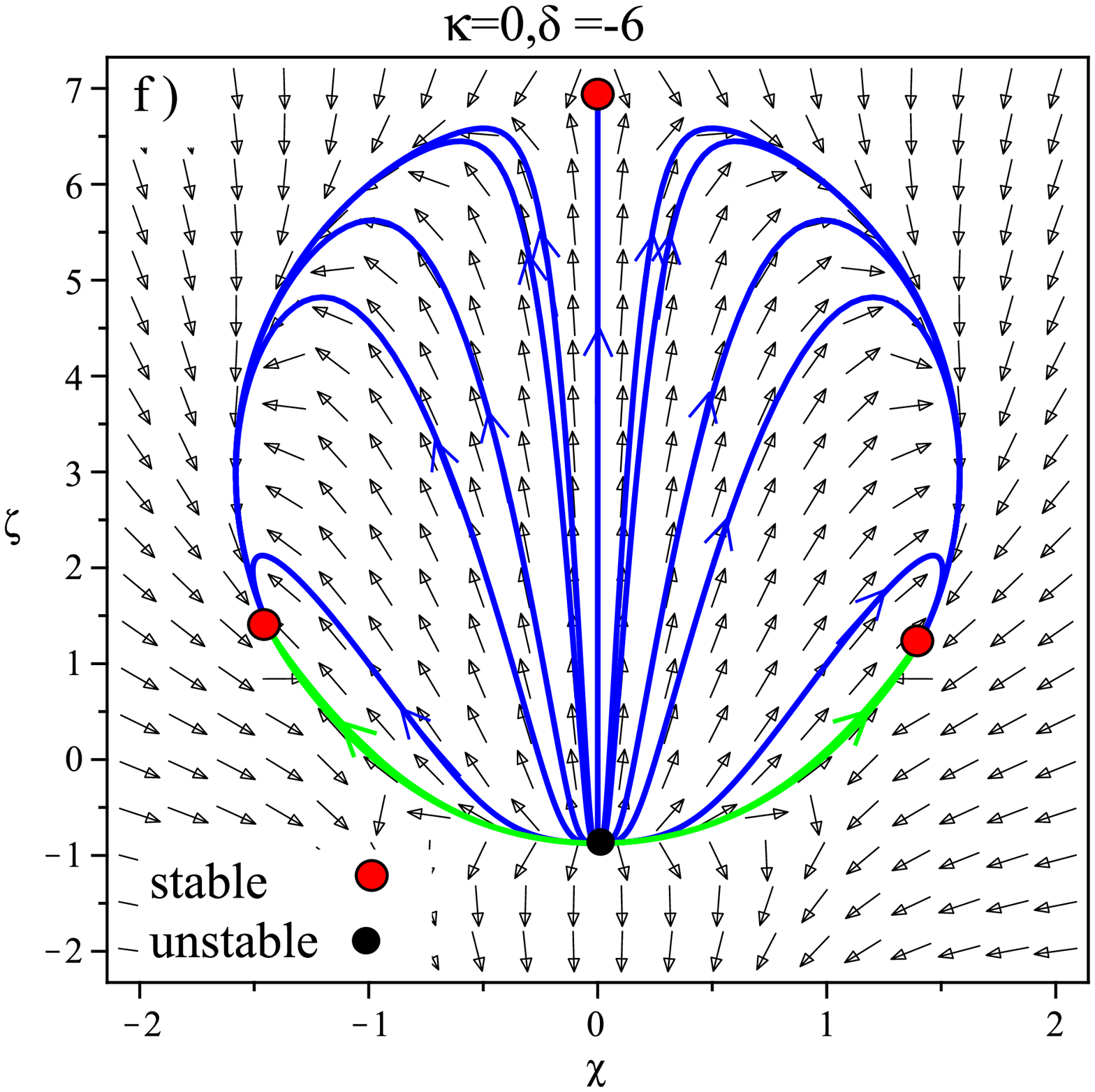}\hspace{0.1 cm}\\
Fig. 5:  The attractor property of the dynamical system in
the phase plane for the case $\gamma=-1$.\\ The plots a) to f) are respectively examples of the critical points $P4, P5, P6$ and $P7$.\\
\end{tabular*}\\

\section{Cosmological parameters: EoS parameter, Statefinders and Distance Modulus}

In order to understand the behavior of the universe and its dynamics we need to study the cosmological parameters such as EoS parameter, deceleration parameter and statefinder. We also need to verify our model with the current observational data  which in here we test the distance modulus. Some of these parameters like EoS and deceleration parameters or even distance modulus analytically and/or numerically have been investigated by many authors for variety of cosmological models. Utilizing stability analysis enables us to have a deeper understanding of the problem and best fit the model with the observational data. For our model the effective EoS and deceleration parameters are defined by $q=-1-\frac{\dot{H}}{H^{2}}$ and  $\omega_{eff}=-1-\frac{2}{3}\frac{\dot{H}}{H^{2}}$ where $\frac{\dot{H}}{H^{2}}$ in terms of new  dynamical variables is given by equation (\ref{hhdot}). The statefinder parameters are another set of parameters that are studied in here. These parameters are defined by $r=\frac{\ddot{H}}{H^{3}}-3q-2$ and $s=\frac{r-1}{3(q-\frac{1}{2})}$ where  $q$ already defined in terms of new dynamical variables and we have $\frac{\ddot{H}}{H^{3}}=\frac{d}{dN}(\frac{\dot{H}}{H^2})+2(\frac{\dot{H}}{H})^2$.

In tables IV to VI for three different scenarios, $\gamma=0, 1/3, -1$, the above cosmological parameters are calculated as:\\

\begin{table}[ht]
\caption{Properties of the critical points for $\gamma=0$ } 
\centering 
\begin{tabular}{c c c c c c } 
\hline\hline 
points  & q & $\omega_{eff}$ & r & s & p\\ [3ex] 
P1  &$ 5+\sqrt{15}$ & $ 3+\frac{2}{3}\sqrt{15}$ &$\frac{203}{2}+\frac{51}{2}\sqrt{15}$ & $\frac{17\sqrt{15}+67}{2\sqrt{15}+9}$
& $\frac{1}{6+\sqrt{15}}$ \\ 
P2  &$ 5-\sqrt{15}$ & $ 3-\frac{2}{3}\sqrt{15}$ &$\frac{203}{2}-\frac{51}{2}\sqrt{15}$ & $\frac{17\sqrt{15}-67}{2\sqrt{15}-9}$ & $\frac{1}{6-\sqrt{15}}$  \\
P3   &$\frac{\delta^2-4\delta-1}{3+\delta}$ & $\frac{2\delta^2-9\delta-5}{9+3\delta}$&$ \frac{B}{(\delta+3)^{3}}$ & $C $& $\frac{\delta^2-3\delta+2}{\delta+3}$\\
P4,5 &$\frac{\kappa^2-4\kappa+24}{32+4\kappa} $& $\frac{\kappa^2-6\kappa+8}{48+6\kappa}$ &$\frac{\kappa^{4}-8\kappa^{3}+76\kappa^2-320\kappa+1184}{8(\kappa+8)^{2}}$ & $\frac{\kappa^{3}-6\kappa^2+56\kappa-336}{6\kappa^{2}+24\kappa-192}$ & $\frac{32+4\kappa}{\kappa^{2}+56}$ \\
P6,7 & $\frac{\kappa^2-36\delta\kappa+56\delta^2-16\kappa-176\delta-72-3\delta\kappa^2+12\delta^2\kappa}{5(-\kappa+4\delta)^2}$ &
  $\omega_{6,7} $&$r_{6,7}$& $s_{6,7} $ & $\frac{1}{q+1}$ \\
 [1ex] 
\hline 
\end{tabular}
\label{table:3} 
\end{table}
 where \\$ \omega_{6,7}=\frac{-3\kappa^2-32\delta\kappa+32\delta^2-32\kappa-352\delta-144-6\delta\kappa^2+24\delta^2\kappa)}{15(-\kappa+4\delta)^2}$,\\
$r_{6,7}=\frac{ (2160-8144\delta^2+1192\delta\kappa-576\delta^3+76\kappa^2+430\delta\kappa^2-2480\delta^2\kappa-11\kappa^3+3\delta \kappa^3-204\delta^2\kappa^2+768\delta^3\kappa-168\kappa+16512\delta)}{5(-\kappa+4\delta)^3}$\\
$s_{6,7}=\frac{2(2160-8144\delta^2+1192\delta\kappa-896\delta^3+76\kappa^2+370\delta\kappa^2-2240\delta^2\kappa-6\kappa^3+3\delta\kappa^3-204\delta^2\kappa^2
+768\delta^3\kappa-168\kappa+16512\delta)}{3(-\kappa+4\delta)(-3\kappa^2-32\delta\kappa+32\delta^2-32\kappa-352\delta-144-6\delta\kappa^2+24\delta^2\kappa)}.$

\begin{table}[ht]
\caption{Properties of the critical points for $\gamma=\frac{1}{3}$ } 
\centering 
\begin{tabular}{c c c c c c } 
\hline\hline 
points  & q & $\omega_{eff}$ & r & s & p\\ [3ex] 
\hline 
P1  &$ 5+\sqrt{15}$ & $ 3+\frac{2}{3}\sqrt{15}$ &$\frac{203}{2}+\frac{51}{2}\sqrt{15}$ & $\frac{17\sqrt{15}+67}{2\sqrt{15}+9}$
& $\frac{1}{6+\sqrt{15}}$ \\ 
P2  &$ 5-\sqrt{15}$ & $ 3-\frac{2}{3}\sqrt{15}$ &$\frac{203}{2}-\frac{51}{2}\sqrt{15}$ & $\frac{17\sqrt{15}-67}{2\sqrt{15}-9}$ & $\frac{1}{6-\sqrt{15}}$  \\
P3   &$\frac{\delta^2-4\delta-1}{3+\delta}$ & $\frac{2\delta^2-9\delta-5}{9+3\delta}$&$ \frac{D}{(\delta+3)^{3}}$ & $E $& $\frac{\delta^2-3\delta+2}{\delta+3}$\\
P4,5 &1 & $\frac{1}{3}$ &3 & $\frac{4}{3}$ & $\frac{1}{2}$ \\
P6,7 & $\frac{\delta-2}{\delta}$ &  $\frac{\delta-4}{3\delta} $&$\frac{(-46\delta+76+3\delta^2)}{\delta^2}$& $\frac{4}{3}\frac{(-23\delta+38+\delta^2)}{\delta(\delta-4)}
$ & $\frac{\delta}{2\delta-2}$ \\
 [1ex] 
\hline 
\end{tabular}
\label{table:3} 
\end{table}

where $D=9-364\delta+10\delta^5+315\delta^2+56\delta^3-74\delta^4$,
and  $E=\frac{2(\delta-2)(-53\delta^2+200\delta+9-54\delta^3+10\delta^4)}{3(-3\delta^2-32\delta-15+2\delta^3)(\delta+3)}$.\\
\begin{table}[ht]
\caption{Properties of the critical points for $\gamma=-1$ } 
\centering 
\begin{tabular}{c c c c c c } 
\hline\hline 
points & q & $\omega_{eff}$ & r & s & p \\ [3ex] 
\hline 
P1  &$ 5+\sqrt{15}$ & $ 3+\frac{2}{3}\sqrt{15}$ &$\frac{203}{2}+\frac{51}{2}\sqrt{15}$ & $\frac{17\sqrt{15}+67}{2\sqrt{15}+9}$
 & $\frac{1}{6+\sqrt{15}}$ \\ 
P2  &$ 5-\sqrt{15}$ & $ 3-\frac{2}{3}\sqrt{15}$ &$\frac{203}{2}-\frac{51}{2}\sqrt{15}$ & $\frac{17\sqrt{15}-67}{2\sqrt{15}-9}$ & $\frac{1}{6-\sqrt{15}}$  \\
P3   &$\frac{\delta^2-4\delta-1}{3+\delta}$ & $\frac{2\delta^2-9\delta-5}{9+3\delta}$&$ \frac{D}{(\delta+3)^{3}}$ & $E $& $\frac{3+\delta}{\delta^2-3\delta+2}$\\
P4,5 & -1 &  -1&1& 0 & $\infty$ \\
  P6,7 & $-\frac{(-7\kappa^2+13\kappa+3-\kappa^3+8\kappa\delta)}{(\kappa+3)^2}$ &$ -\frac{1}{3}\frac{(-13\kappa^2+32\kappa+15-2\kappa^3+16\kappa\delta)}{(\kappa+3)^2}$&F&
 G & $\frac{-(\kappa+3)^2}{(-8\kappa^2+7\kappa-6-\kappa^3+8\kappa\delta)}$ \\
 [2ex] 
\hline 
\end{tabular}
\label{table:5} 
\end{table}\\
where $F=\frac{-(-9-112\kappa^2\delta-6\kappa^4+264\kappa\delta-160\kappa\delta^2+80\kappa^2\delta^2+32\kappa^3-2\kappa^5+12\kappa-99\kappa^2)}{(\kappa+3)^3}$,\\
and $G=\frac{\frac{2}{3}(18-112\kappa^2\delta-6\kappa^4+264\kappa\delta-160\kappa\delta^2+80\kappa^2\delta^2+33\kappa^3
-2\kappa^5+39\kappa-90\kappa^2)}{(-13\kappa^2+32\kappa+15-2\kappa^3+16\kappa\delta)(\kappa+3)}$.

In the above tables $p$ is the power of time $t$ in $a(t)\varpropto t^p$ obtained in terms of the dynamical variables. For $\gamma=0$, from tables I and IV, the critical points P1 and P2 are independent of the stability parameters $\kappa$ and $\delta$, having numerical values. These critical points show a decelerating universe with positive effective EoS parameter. On the other hand, the critical point P3 is stable under some conditions on $\delta$ while independent of $\kappa$ while the points P4 to P7 are both $\delta$ and $\kappa$ dependent. All these critical points can be regarded as attractors for all the trajectories in the statefinder phase plane if satisfy the stability condition. In case of $\gamma=1/3$, as can be seen from tables II and V, the unstable critical points P1, P2, P4 and P5 are independent of the stability parameters $\kappa$ and $\delta$, given the numerical values. On the other hand, the critical points P3 , P6 and P7 are stable under some conditions on $\delta$ while independent of $\kappa$ and so can be regarded as attractors for all the trajectories in the statefinder phase plane. For $\gamma=-1$, from tables III and VI, while the critical points P1, P2, P4 and P5 are dependent of the conditions on $\kappa$ and $\delta$, their physical parameters have numerical values. On the other hand the critical points P3 , P6 and P7 are independent of $\kappa$. All these points can be regarded as attractors for the trajectories in the statefinder phase plane.

We can discuss the dynamics of the EoS parameter and deceleration parameter in Fig. 6) in the case of $\gamma=0$. As is shown, for a variety of stability parameters, the evolving effective EoS parameter, $\omega_{eff}$, and the deceleration parameter $q$ are numerically computed. From the graph, the EoS parameters for the stability parameters $\delta=6$, $\kappa=3$ (green curve) do not hit the phantom divide line , while for $\delta=7$, $\kappa=1$ and $\delta=6$,$\kappa=3$ ( red and blue curves) the trajectories cross the line. It also shows a symmetrical behavior in the parameters about $ln (a)=0.15$ ( $z=-0.139$). Moreover, the deceleration parameter, $q$ at the present time is negative and between about $-0.9$ and $-2.1$ (an accelerating universe) for the given stability parameters. For both blue and red curves, the crossing occurs at about $ln (a)=-0.2$ ( $z=0.221$) in the past while in the future at  $ln (a)=0.35$ ($z=-0.3$) for blue curve and
$ln (a)=0.5$ ($z=-0.393$) for red curve. Using the above argument, and from the Fig. 13)left) one can see that all the trajectories crossing phantom line in the interval $z=0-1$ are best fitted with the observational data while the red curve in the interval $z=1-2$ better fits the data. Noting that all the trajectories satisfy the stable critical points P6,7 condition. With the selected stability parameters in blue trajectory, $\omega_{eff}=-1/9$ and in the red and green trajectories $\omega_{eff}=-1/21$ which in both cases it occurs sometimes in the far past and future, as can be seen in the graph.\\

\begin{tabular*}{2.5 cm}{cc}
\includegraphics[scale=.35]{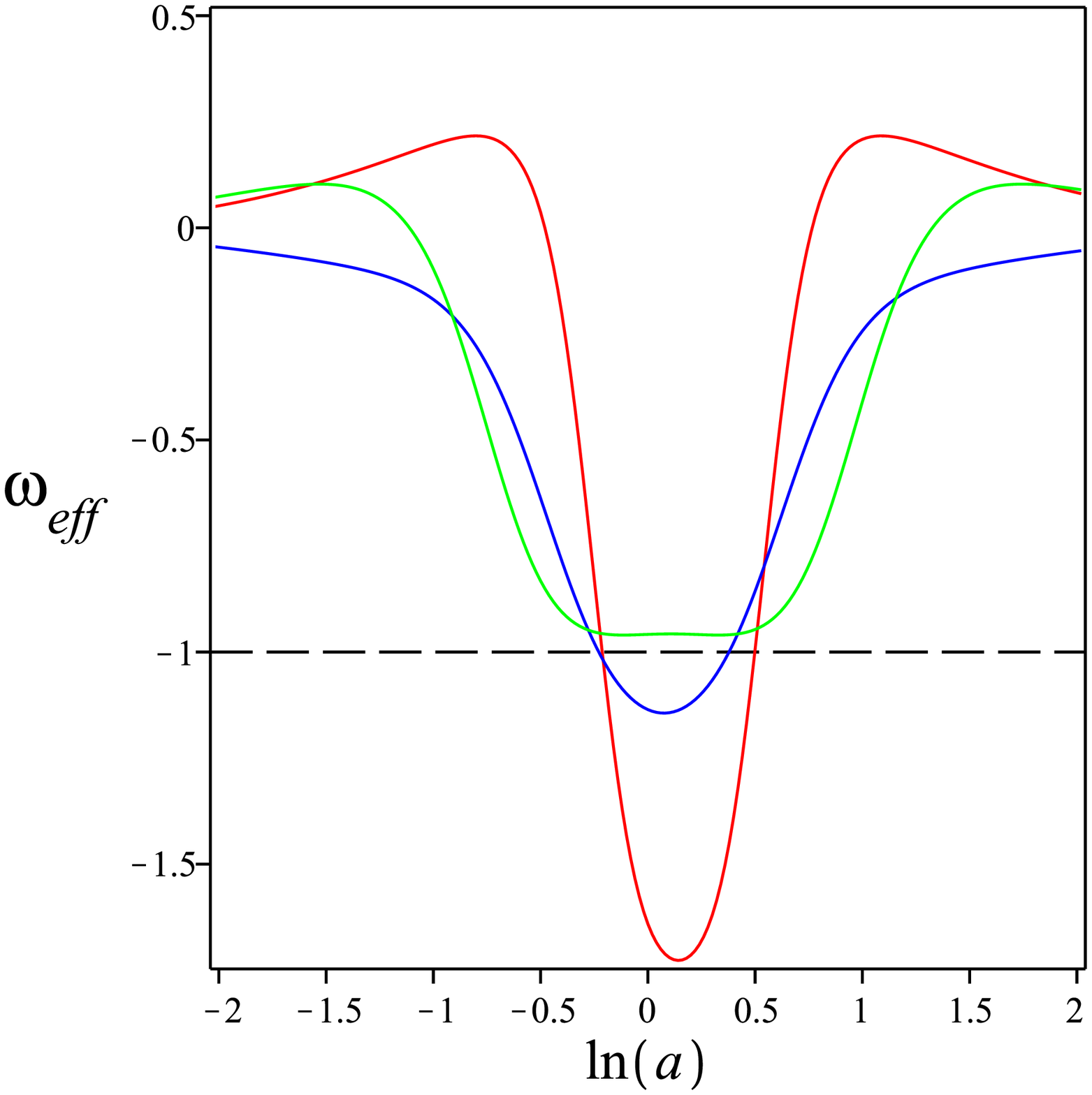}\hspace{0.1 cm}\includegraphics[scale=.35]{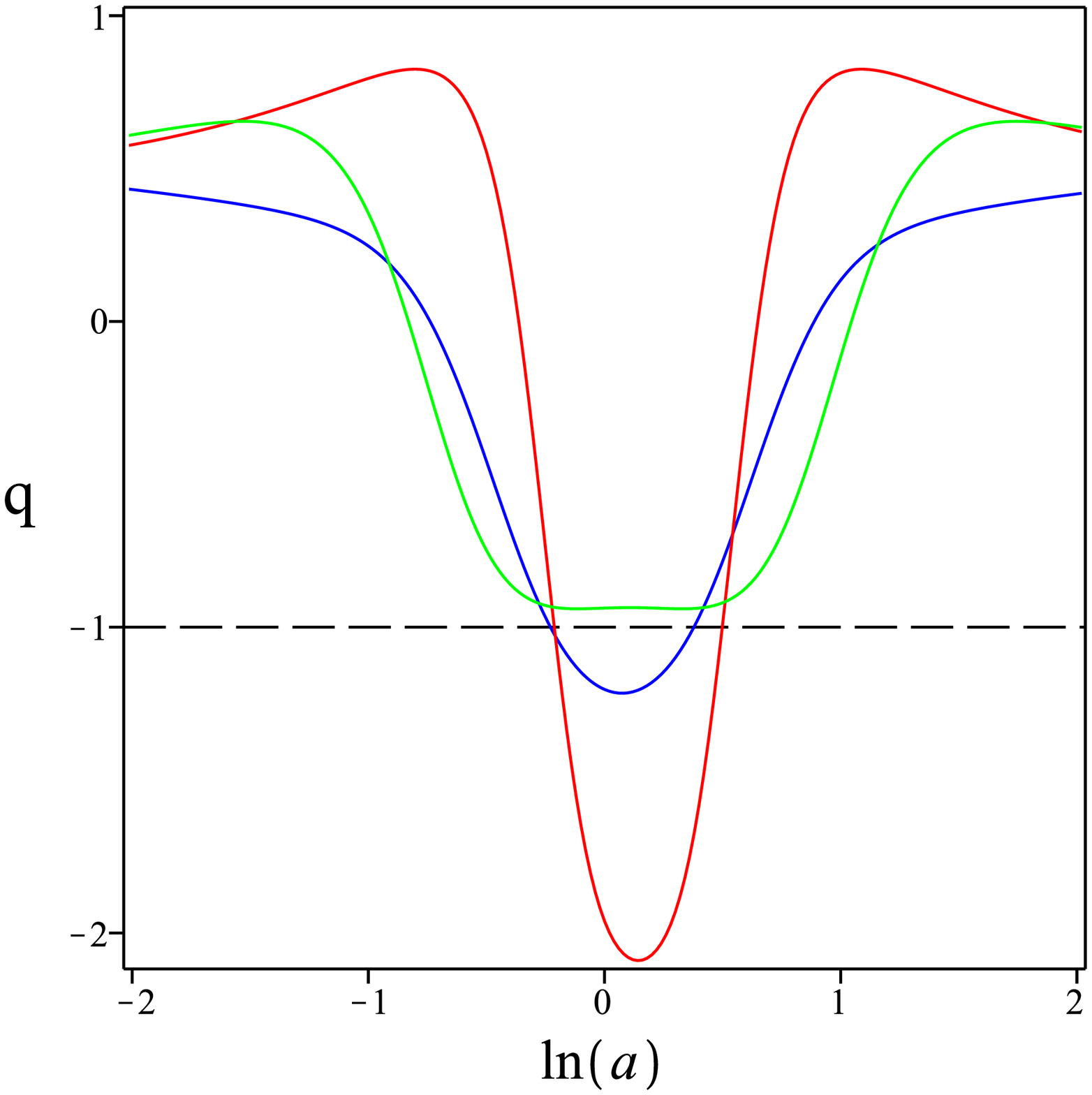}\hspace{0.1 cm}\\
Fig. 6:  The evolution of the effective EoS parameter $\omega_{eff}$ and deceleration parameter $q$\\
 as a function of $z$ for $\gamma=0$. Stability conditions:
 (blue)$\delta=7$,$\kappa=1$-(green $\&$ red)$\delta=6$,$\kappa=3$\\
 I.Cs.:(blue \& green)$\chi(0)=0.1$, $\zeta(0)=-0.5$
 (red)$\chi(0)=1/3$, $\zeta(0)=-2/9$\\
\end{tabular*}\\

In Fig. 7), the evolving effective EoS parameter, $\omega_{eff}$, and the deceleration parameter $q$ are shown for $\gamma=1/3$. It can be seen that the EoS parameter for the stability parameter $\delta=4$ (blue curve) does not hit the phantom divide line while for  $\delta=6$ (black curve) crosses the line. It also shows a symmetrical behavior of the curves sometimes in the future about lines $ln (a)=0.2$ ( $z=-0.18$) and  $ln (a)=1.5$ ( $z=-0.78$) respectively for the above two stability parameters. Moreover, the trajectory in the case $\delta=4$ while symmetrically and sinusoidally oscillates, eventually diminishes in far past and future. The graph also shows that the deceleration parameter, $q$, which has a similar behavior to the effective EoS parameter, is at the present time negative and $-0.6$ or $-1.4$ (an accelerating universe) for the given stability parameters. With respect to these two different EoS trajectories and stability analysis, a question arises as to the validity of the solutions since one of them is crossing the phantom line and the other one not. The answer is that the trajectory which best fit the observational data should be considered as physical one. From Fig. 13)middle), as will be discussed, one can see that the difference in the distance modulus is best fitted with the observational data for the trajectory that crosses the phantom line. Noting that both trajectories correspond to the stable critical points, the blue one satisfies the critical points P3)ii) condition, while the black one satisfies the P6,7)i) condition. With the selected stability parameter in blue trajectory, $\omega_{eff}=-0.5$ and in the black trajectory $\omega_{eff}=1/9$ which in both cases it occurs sometimes in the past and future as can be seen in the graph.\\

\begin{tabular*}{2.5 cm}{cc}
\includegraphics[scale=.35]{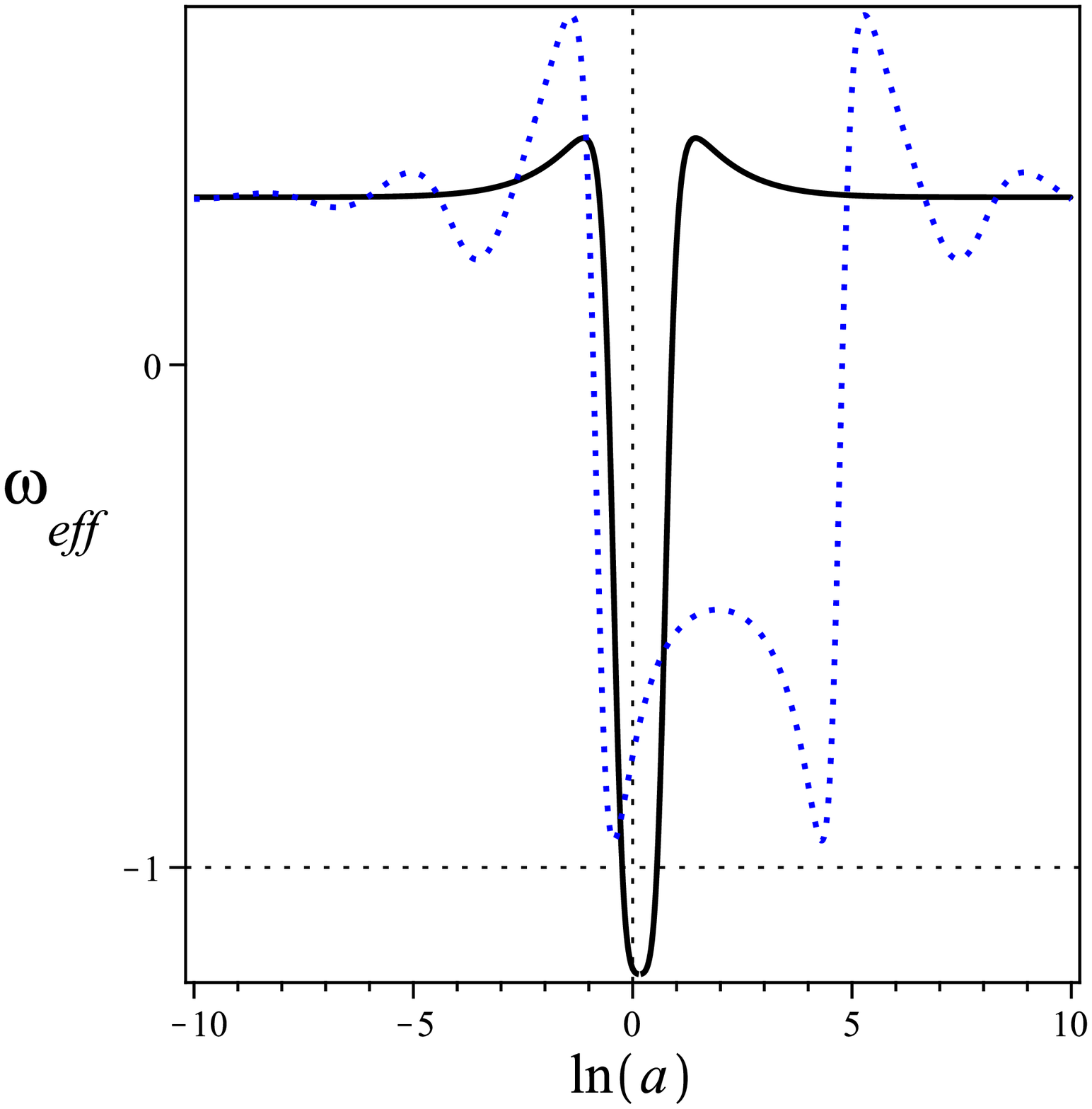}\hspace{0.1 cm}\includegraphics[scale=.35]{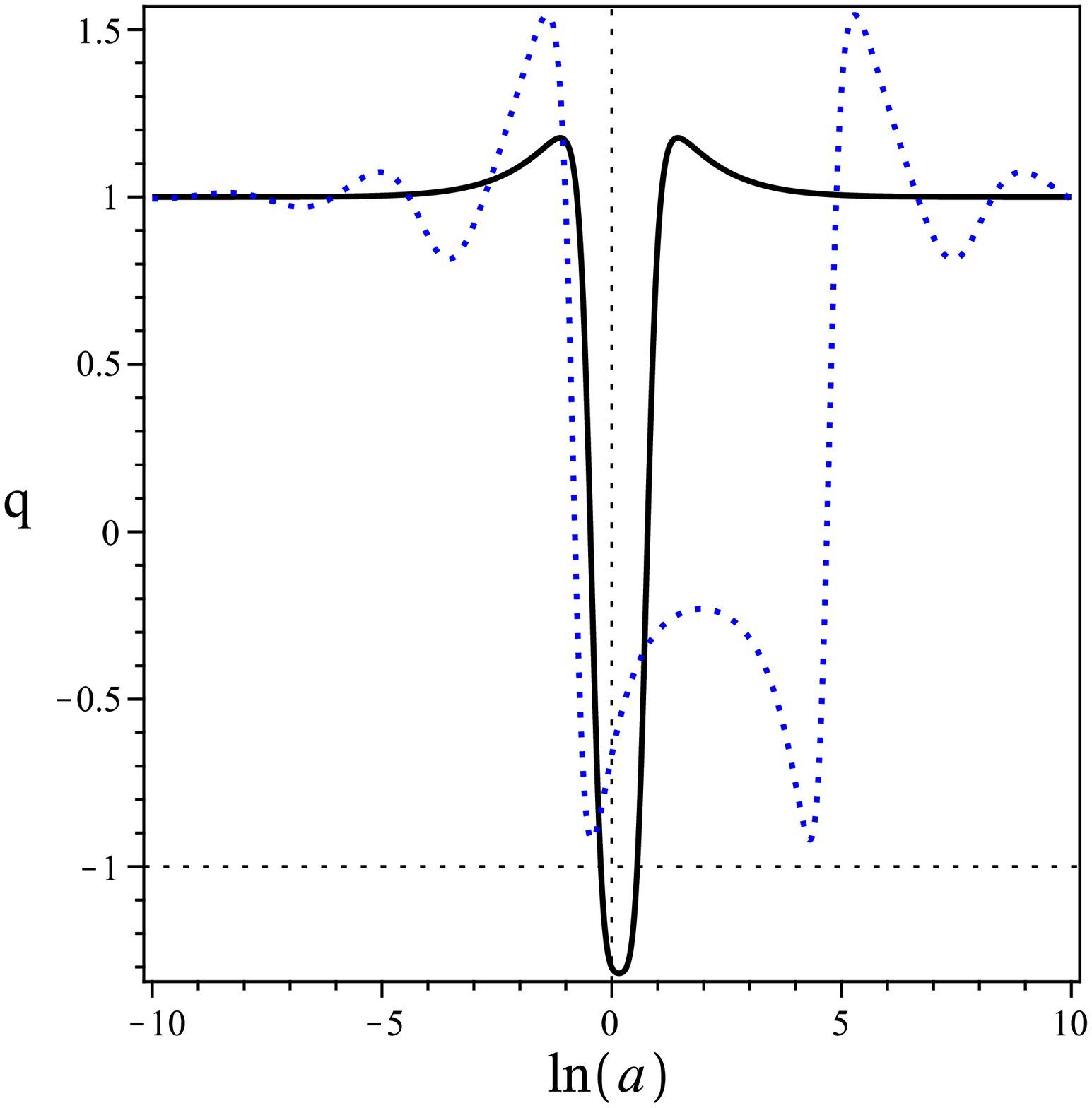}\hspace{0.1 cm}\\
Fig. 7:  The evolution of the effective EoS parameter $\omega_{eff}$ and deceleration parameter $q$\\
 as a function of $ln(a)$ for $\gamma=1/3$
 (blue):$\delta=4$, (black):$\delta$=6 - I.C. $\chi(0)=0.21$, $\zeta(0)=-0.4$\\
\end{tabular*}\\

With the same argument as above, we can discuss the dynamics of the EoS parameter and deceleration parameter in Fig. 8) in the case of $\gamma=-1$. As is shown, for a variety of stability parameters, the evolving effective EoS parameter, $\omega_{eff}$, and the deceleration parameter $q$ are numerically computed. From the graph, the EoS parameters for the stability parameters $\delta=6$, $\kappa=3$ do not hit the phantom divide line (green and black curves), while for $\delta=7$, $\kappa=1$ and $\delta=5$,$\kappa=5$ the trajectories cross the line ( red and blue curves). It also shows a symmetrical behavior in the parameters about $ln (a)=0.1$ ( $z=-0.1$)and $ln (a)=0.2$ ( $z=-0.18$). Moreover, the deceleration parameter, $q$ at the present time is negative and between about $-1$ and $-2.3$ (an accelerating universe) for the given stability parameters. For the stability parameter $\delta=7$, $\kappa=1$ (blue curve), the crossing occurs at $ln (a)=0.3$ ( $z=-0.26$) in the future and  $ln (a)=-0.1$ ( $z=0.11$) in the past.
 For the stability parameter $\delta=5$, $\kappa=5$ (red curve), it occurs at $ln (a)=1.2$ ( $z=-0.7$) in the future and  $ln (a)=-1.2$ ( $z=2.3$) in the past and then again become tangent to the line in the future and past. Using the above argument, and from the Fig. 13)right) one can see the blue trajectory crossing phantom line is best fitted with the observational data in comparison with the other trajectories. Noting that the red trajectory correspond to a critical point that is not stable, the green and black trajectories both with the same stability parameter and different initial conditions satisfy the critical points P6,7)i) condition, while the blue trajectory satisfies the P3)iii) condition.\\

\begin{tabular*}{2.5 cm}{cc}
\includegraphics[scale=.35]{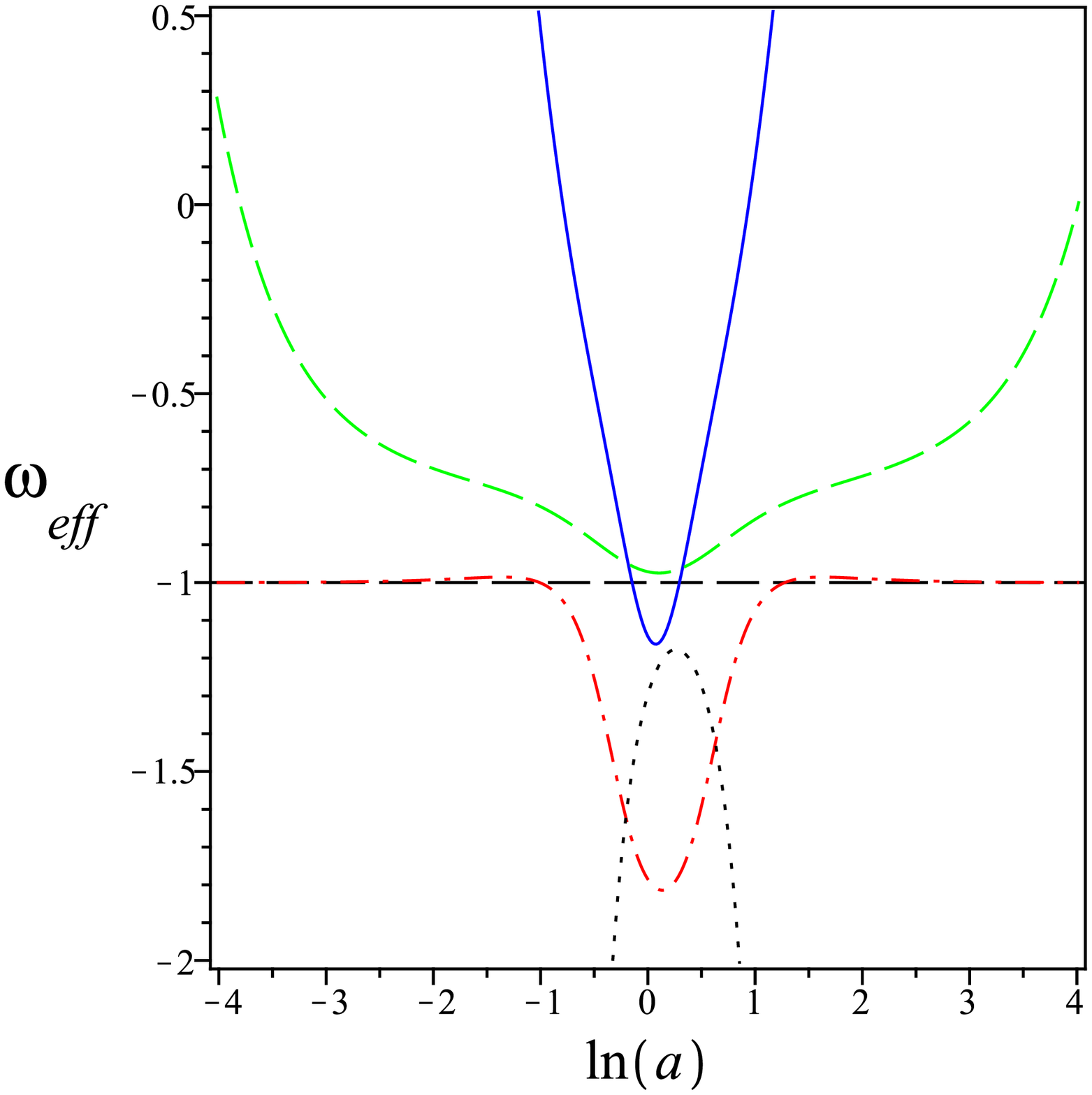}\hspace{0.1 cm}\includegraphics[scale=.35]{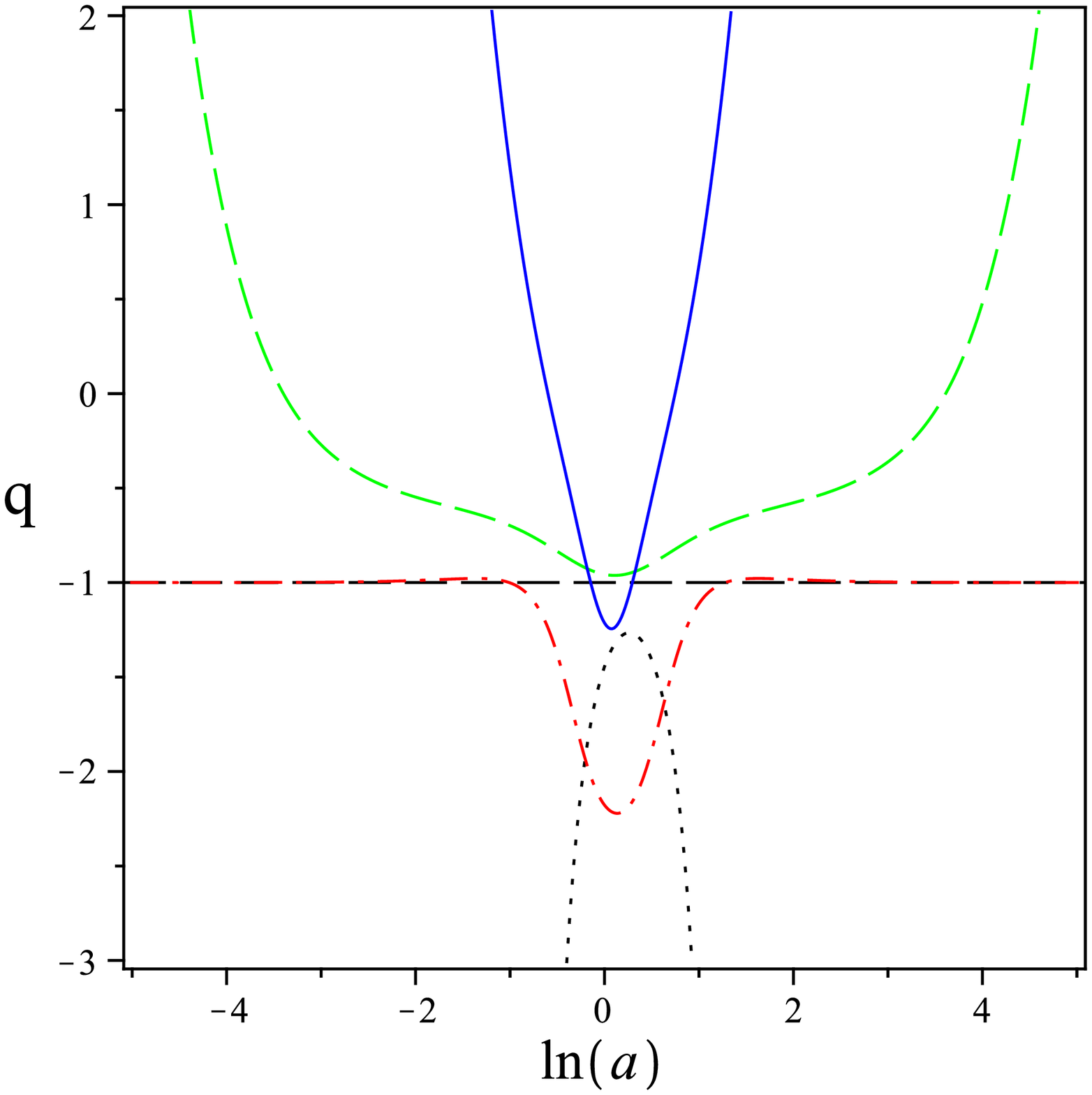}\hspace{0.1 cm}\\
Fig. 8:  The evolution of the effective EoS parameter $\omega_{eff}$ and deceleration parameter $q$\\
 as a function of $ln(a)$ for $\gamma=-1$. Stability conditions:
 (blue)$\delta=7$,$\kappa=1$\\-(red)$\delta=5$,$\kappa=5$-(green $\&$ black)$\delta=6$,$\kappa=3$
 I.Cs.:(blue \& green)$\chi(0)=0.1$, $\zeta(0)=-0.5$\\
 (red)$\chi(0)=1/3$, $\zeta(0)=-2/9$-(black)$\chi(0)=3/10$, $\zeta(0)=-1/3$\\
\end{tabular*}\\

As a complementarity, Fig. 9) shows statefinder diagrams for $\{s,q\}$ and $\{r,q\}$ evolutionary
trajectories for $\gamma=0$. From this figure, we clearly see the position of the current values of the statefinder parameters for the two given stability conditions with respect to the LCDM, SCDM and SS states. The trajectories commence evolving from the same state and end their evolution to the same common state.

\begin{tabular*}{2.5 cm}{cc}
\includegraphics[scale=.35]{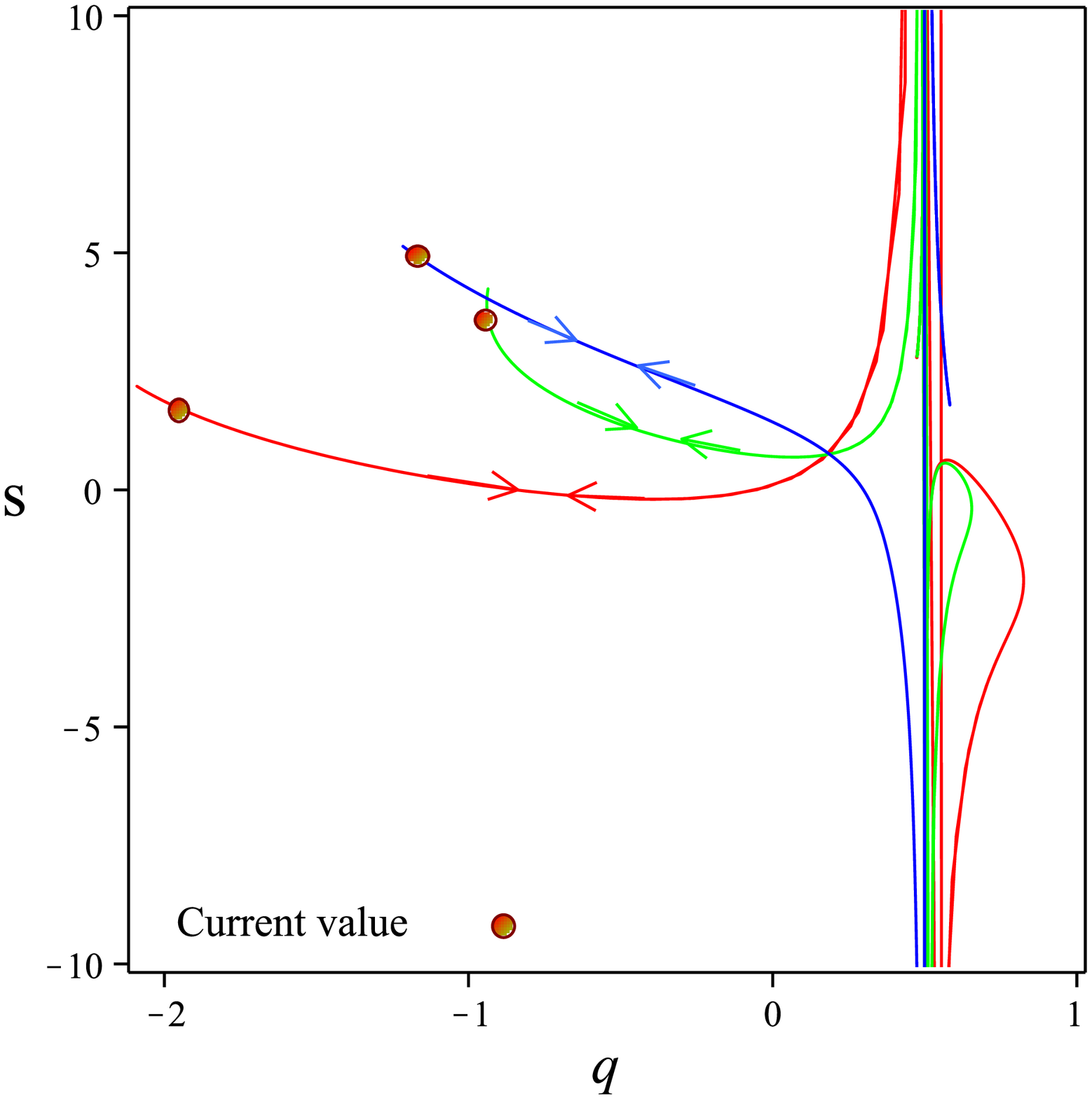}\hspace{0.1 cm}\includegraphics[scale=.35]{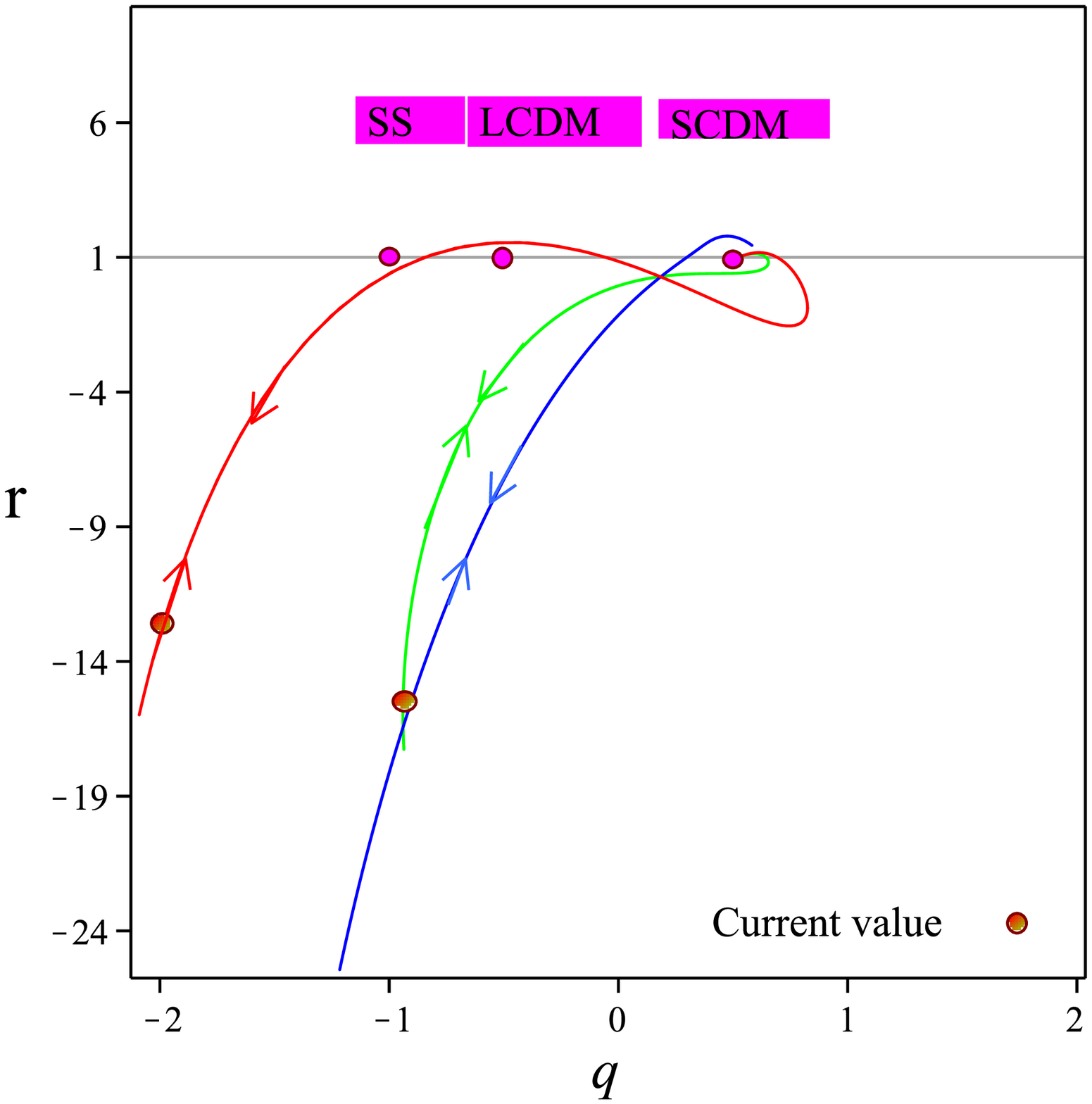}\hspace{0.1 cm}\\
Fig. 9:  Trajectories in the statefinder plane $\{r, q\}$ and $\{s, q\}$ for the\\
model with $\gamma=0$.The arrows show the direction of the time evolution.\\
 Stability conditions:(blue)$\delta=7$,$\kappa=1$-(green $\&$ red)$\delta=6$,$\kappa=3$\\
 I.Cs. (blue \& green)$\chi(0)=0.1$, $\zeta(0)=-0.5$
 (red)$\chi(0)=1/3$, $\zeta(0)=-2/9$\\
\end{tabular*}\\

Also, Fig. 10) shows statefinder diagrams for $\{s,q\}$ and $\{r,q\}$ evolutionary
trajectories for $\gamma=1/3$. From this figure, we clearly see the position of the current values of the statefinder parameters for the two given stability conditions with respect to the LCDM, SCDM and SS states. The graph also shows that in both cases the trajectories evolve to an unstable critical point. They commence evolving from the same state and end their evolution to the same common state.\\

\begin{tabular*}{2.5 cm}{cc}
\includegraphics[scale=.35]{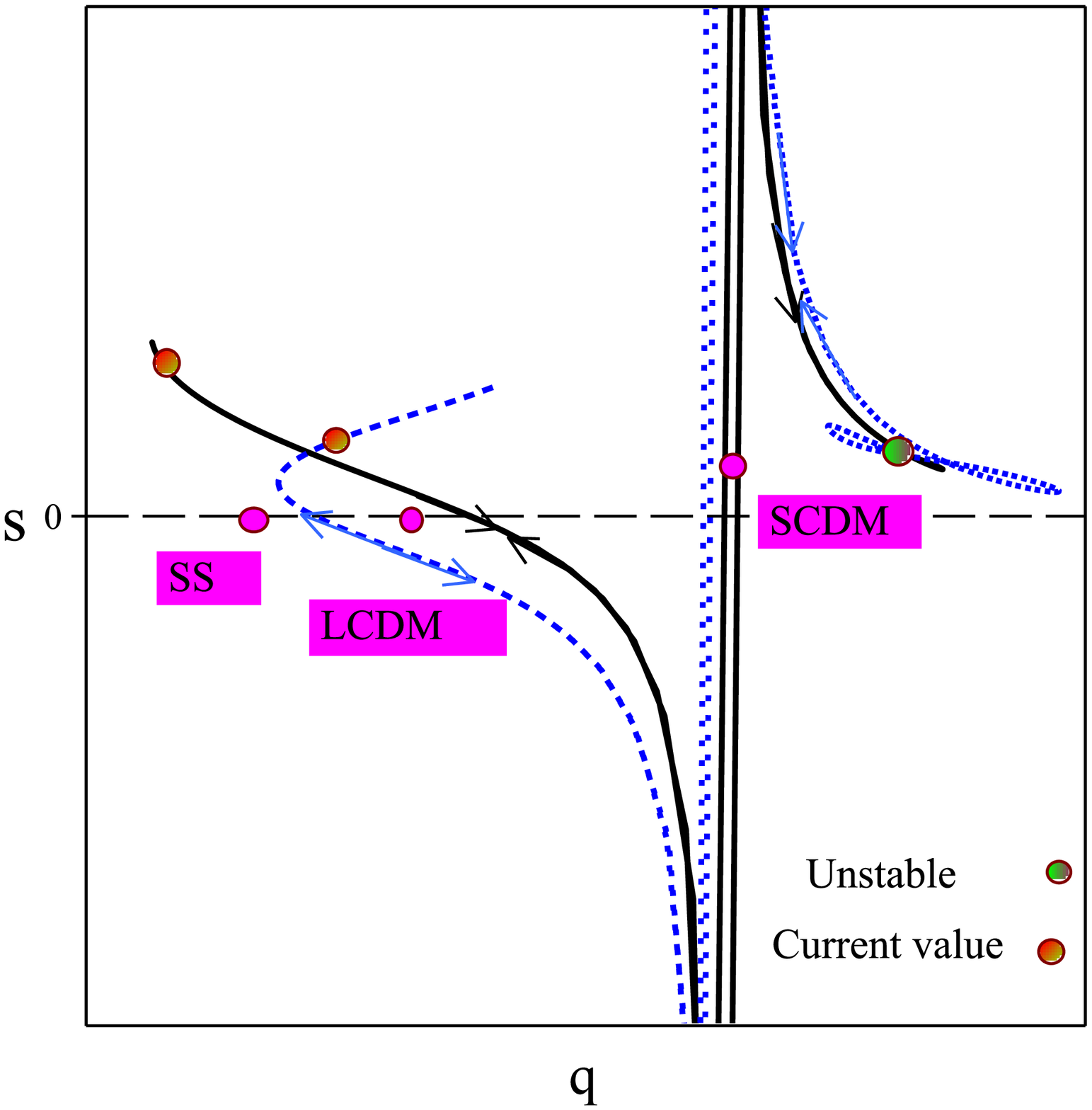}\hspace{0.1 cm}\includegraphics[scale=.35]{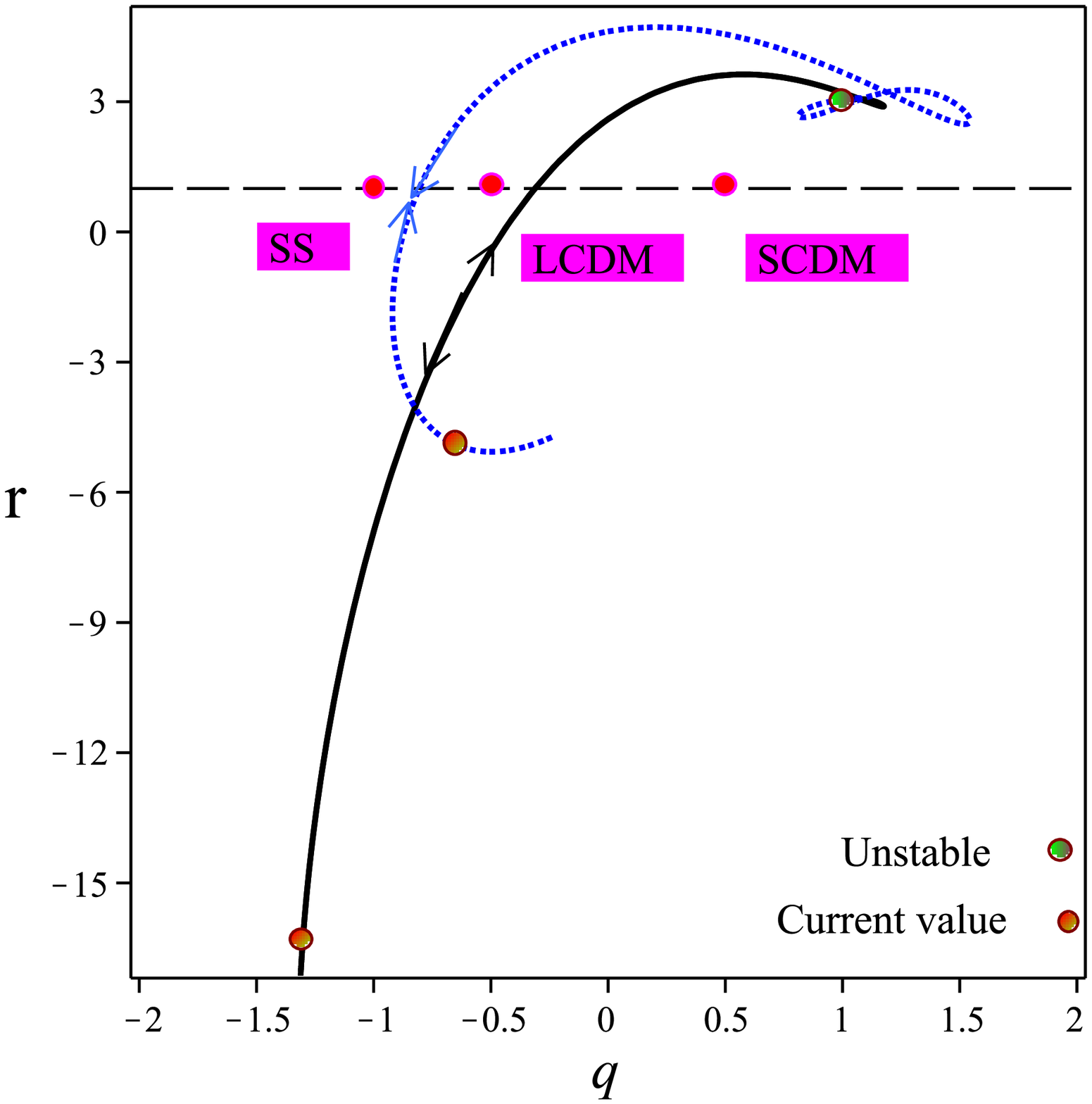}\hspace{0.1 cm}\\
Fig. 10:  Trajectories in the statefinder plane $\{r, q\}$ and $\{s, q\}$ for the\\
model with $\gamma=1/3$.The arrows show the direction of the time evolution.\\
Stability conditions:(blue)$\delta=4$-(black)$\delta=6$.
 I.Cs. $\chi(0)=0.21$, $\zeta(0)=-0.4$
\end{tabular*}\\

in the case of $\gamma=-1$, Fig. 11) shows statefinder diagrams for $\{s,q\}$ and $\{r,q\}$  evolutionary
trajectories. From this figure, the position of the current values of the statefinder parameters for the four given stability conditions with respect to the LCDM, SCDM and SS states are shown. Likewise the other cases, the trajectories commence evolving from the same state in the past and end their evolution to the same common state.\\

\begin{tabular*}{2.5 cm}{cc}
\includegraphics[scale=.35]{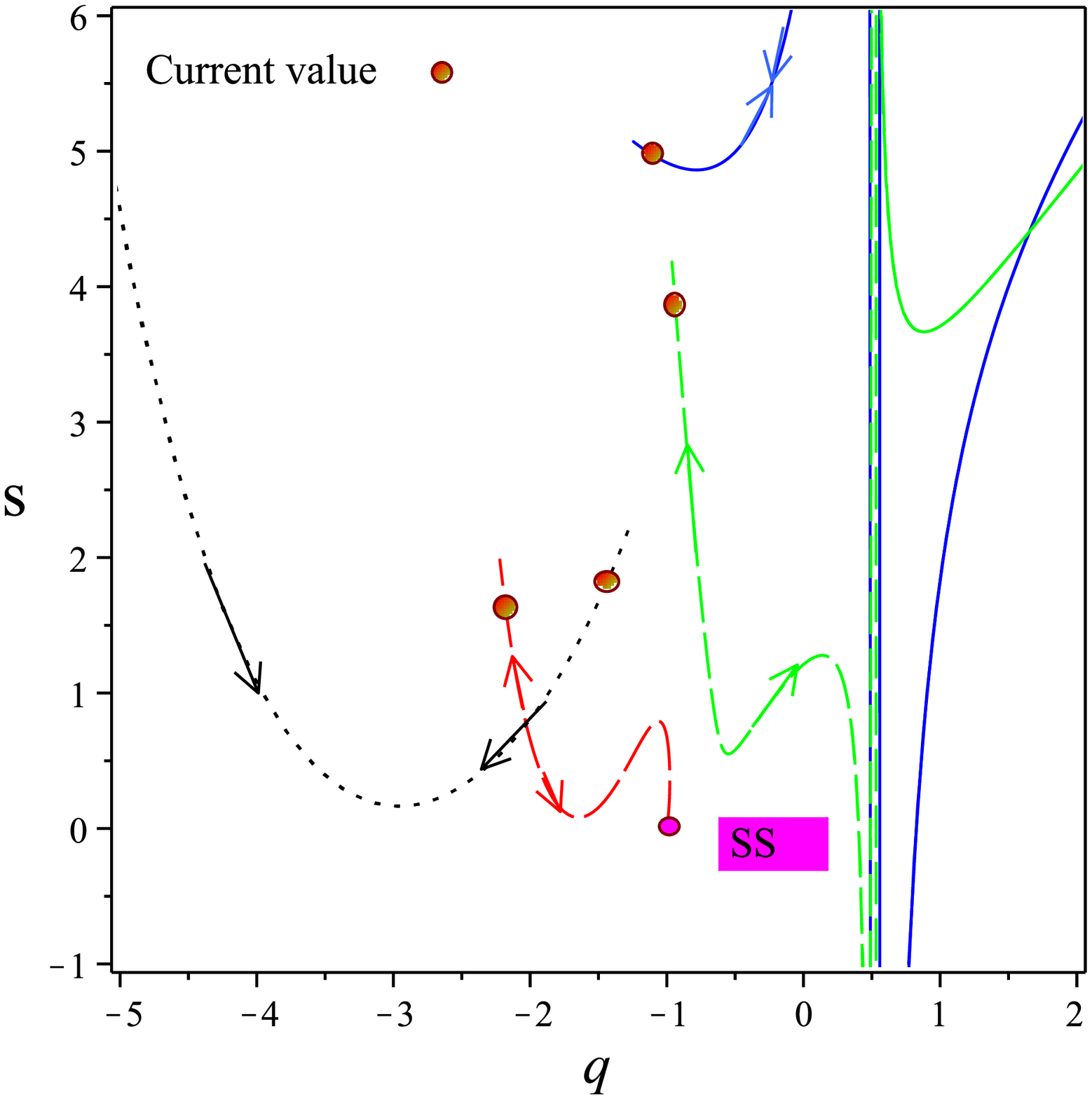}\hspace{0.1 cm}\includegraphics[scale=.35]{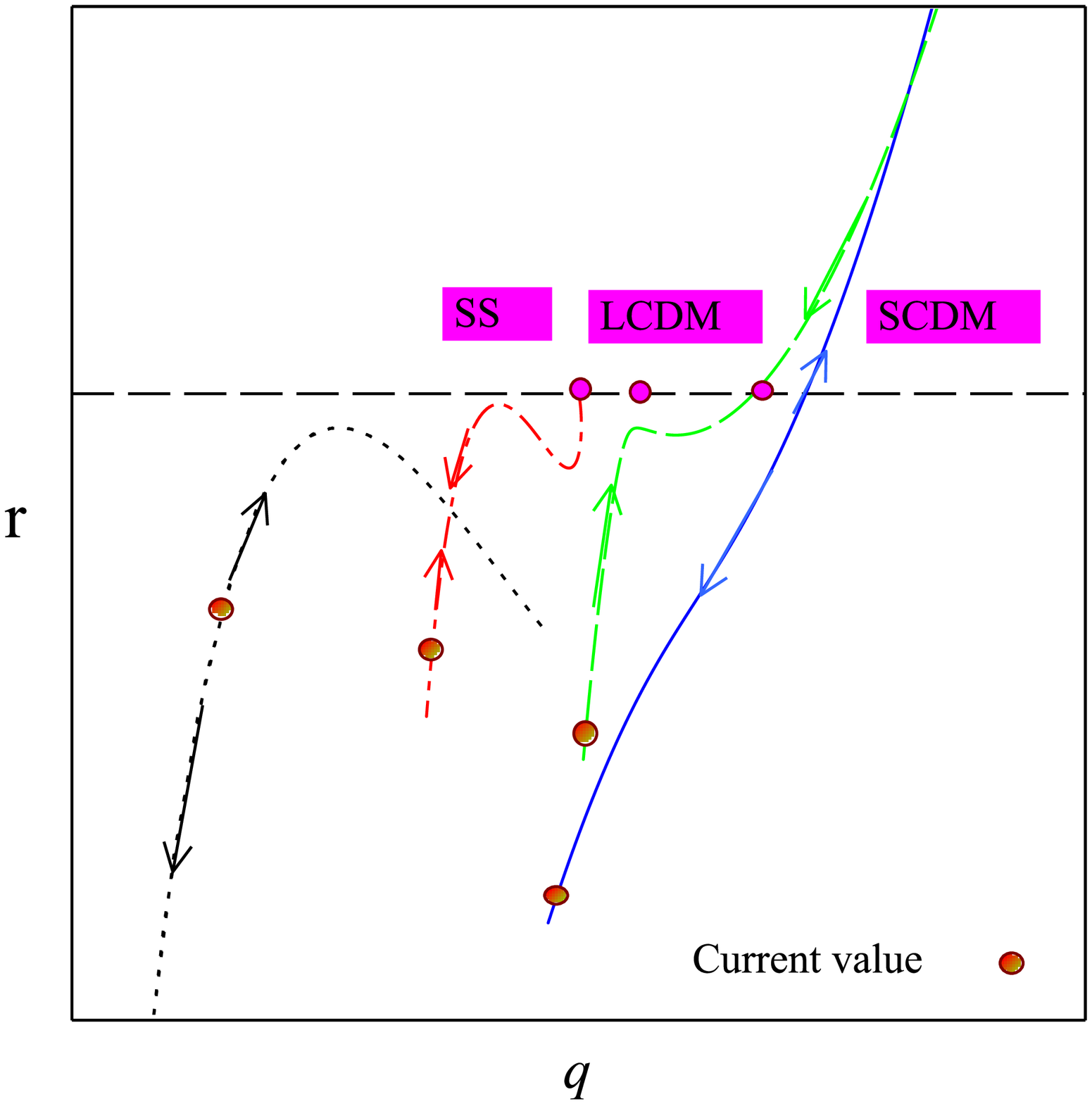}\hspace{0.1 cm}\\
Fig. 11:  Trajectories in the statefinder plane $\{r, q\}$ and $\{s, q\}$ for the model with $\gamma=-1$.\\
The arrows show the direction of the time evolution.
 Stability conditions:(blue)$\delta=7$,$\kappa=1$-\\
 (green)$\delta=6$,$\kappa=3$-(red)$\delta=5$,$\kappa=5$-(black)$\delta=6$,$\kappa=3$
 I.Cs. (blue \& green)$\chi(0)=0.1$, $\zeta(0)=-0.5$\\
 (red)$\chi(0)=1/3$, $\zeta(0)=-2/9$-(black)$\chi(0)=3/10$, $\zeta(0)=-1/3$\\
\end{tabular*}\\

In Fig. 12), the statefinder diagrams — $\{r, s\}$ for $\gamma=0$(left), $\gamma=1/3$(middle) and $\gamma=-1$(right) are shown. Again the current state of the universe for the given stability conditions with respect to the LCDM scenario is presented. The graph shows that in case of $\gamma=1/3$ the trajectories start from an unstable critical point and end to the same unstable point which may be indirect evidence of a cyclic universe predicted by string theory \cite{Habara}. The current values of the trajectories in our model for $\gamma=0, 1/3, -1$ and their distance from LCDM scenario can be
seen explicitly in these diagrams. Note that the trajectories
of our model will pass through LCDM fixed point.
It is of interest to see that the trajectories have different forms
before reaching the LCDM state \cite{Sahni}. This behavior demonstrated that the statefinder can successfully
characterize and differentiate between various DE models.\\

\begin{tabular*}{2.5 cm}{cc}
\includegraphics[scale=.30]{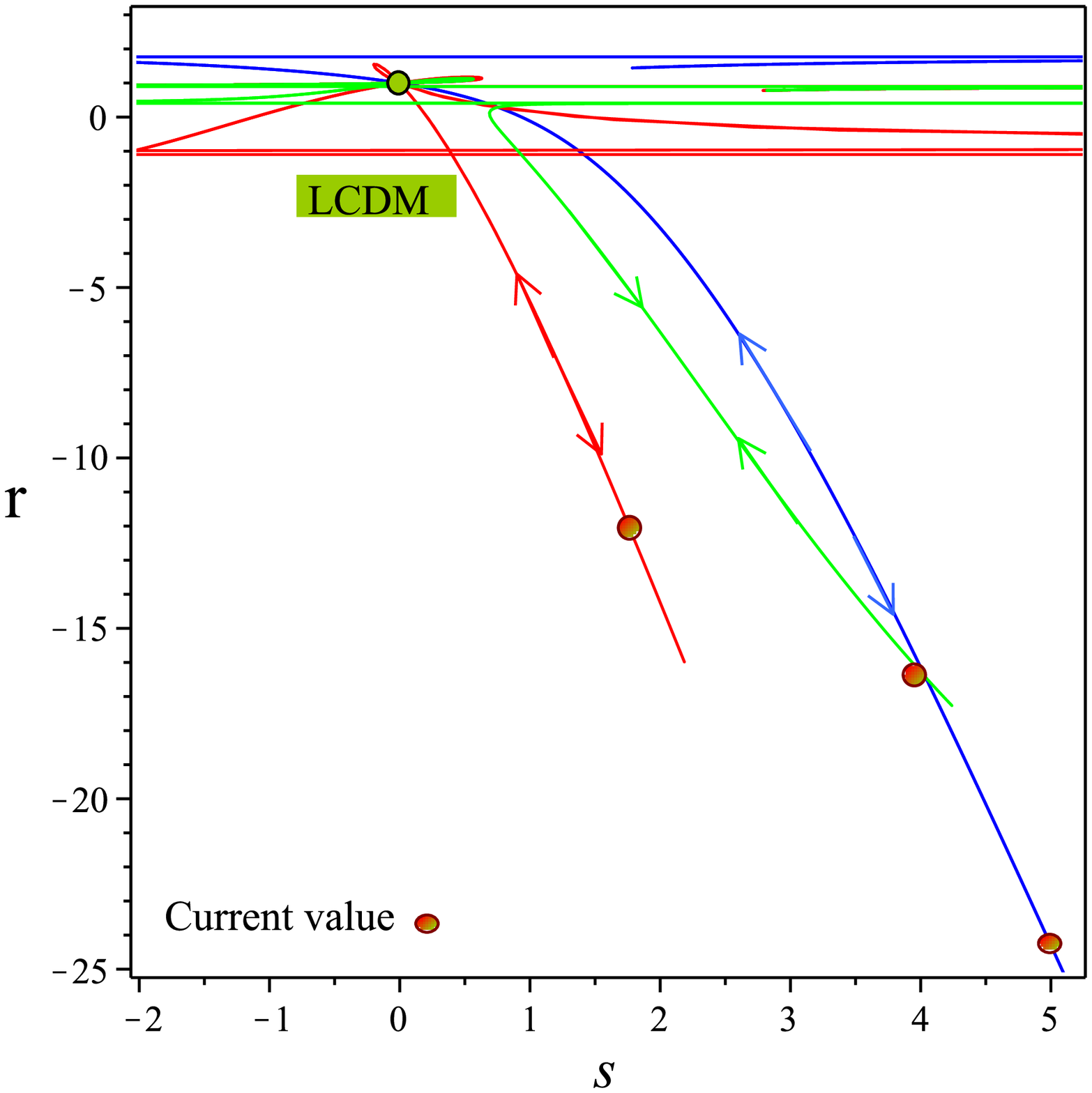}\hspace{0.1 cm}\includegraphics[scale=.30]{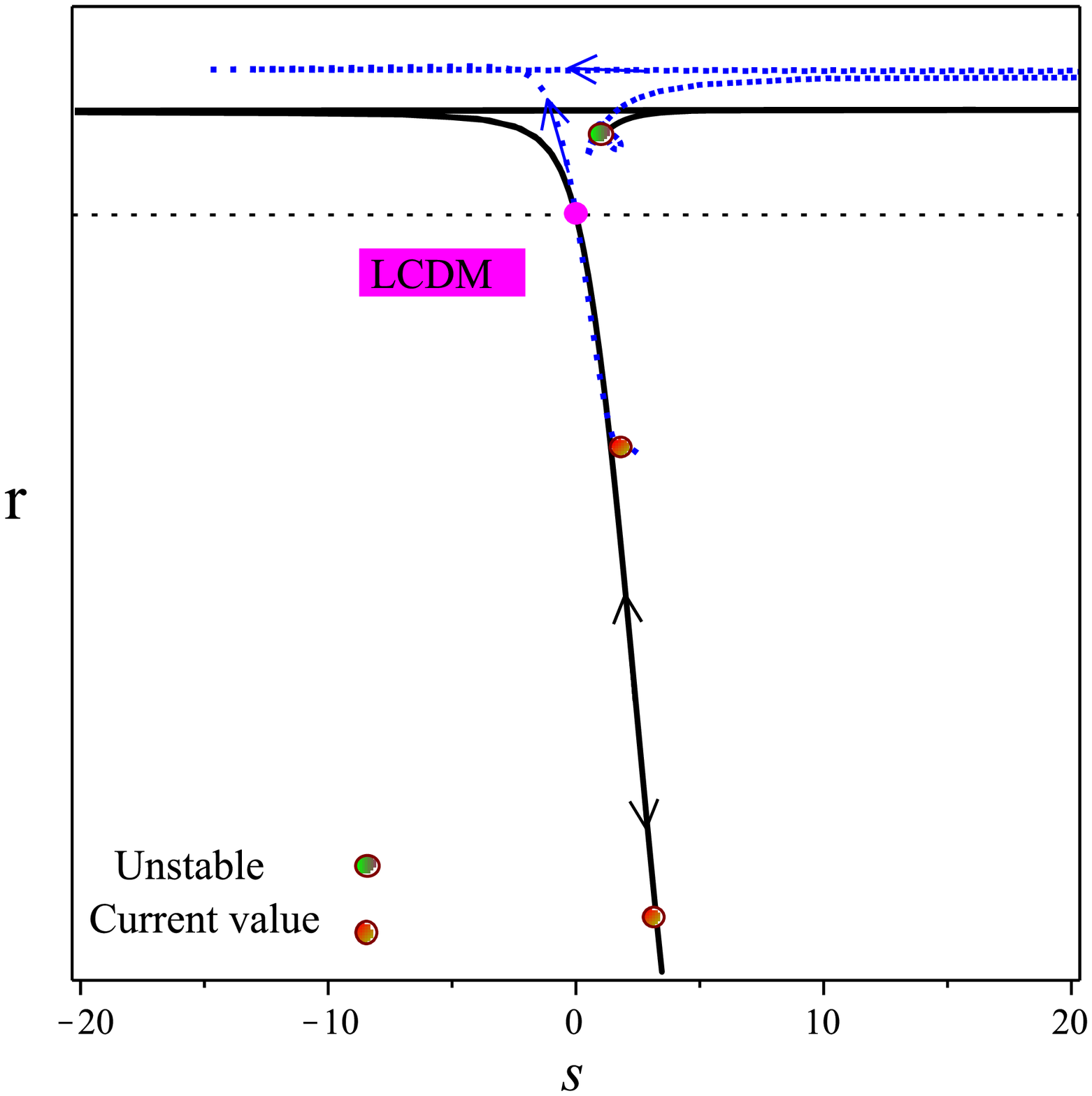}\hspace{0.1 cm}\includegraphics[scale=.30]{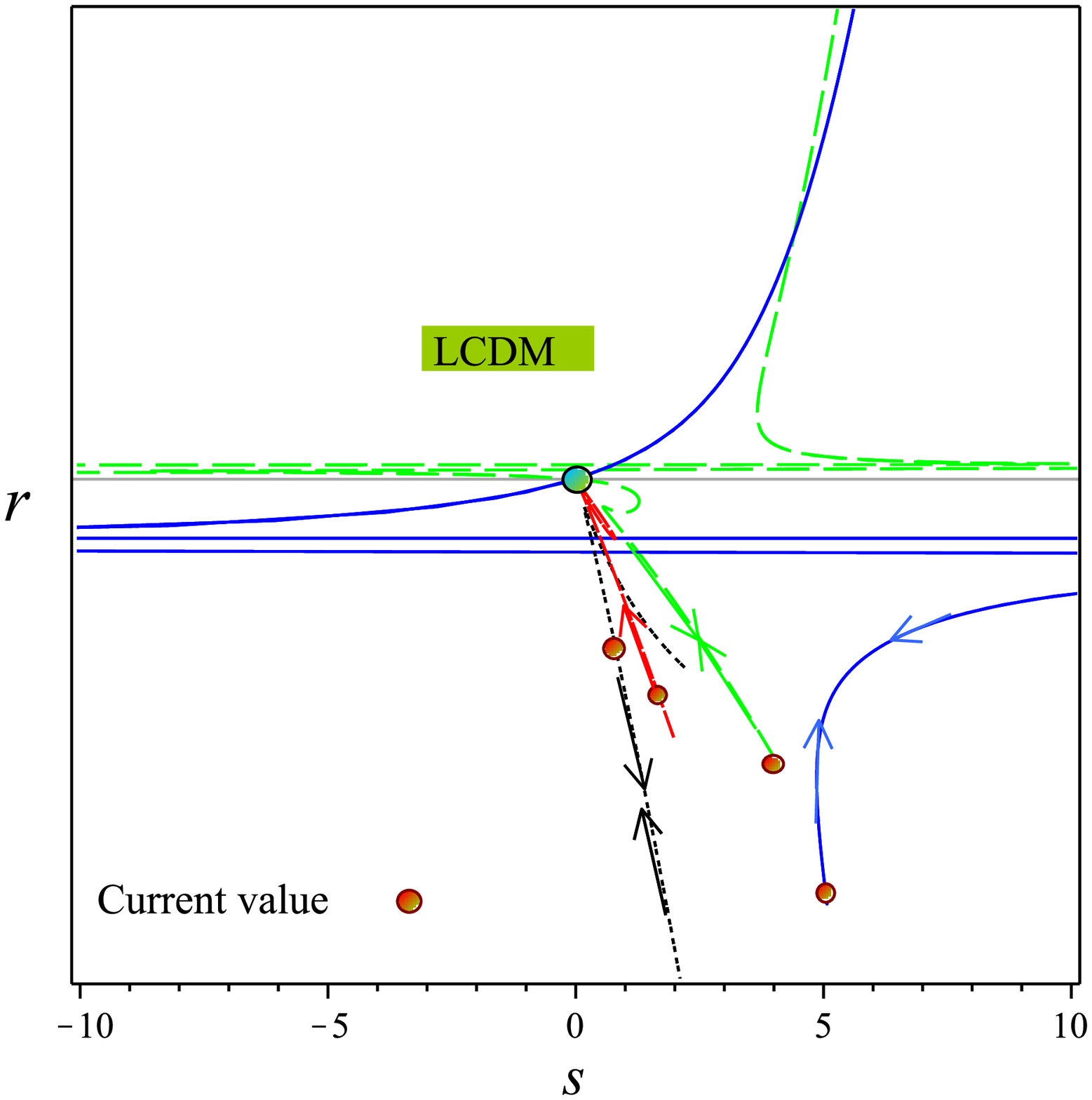}\\
Fig. 12: Trajectories in the statefinder plane $\{r, s\}$ for $\gamma=0$(left), $\gamma=\frac{1}{3}$(middle),\\ $\gamma=-1$(right). 
For $\gamma=0$:(blue)$\delta=7$,$\kappa=1$-(green \&red)$\delta=6$,$\kappa=3$\\
 I.Cs.:(blue \& green)$\chi(0)=0.1$, $\zeta(0)=-0.5$
 (red)$\chi(0)=1/3$, $\zeta(0)=-2/9$. \\For $\gamma=\frac{1}{3}$: (blue)$\delta=4$-(black)$\delta=6$. I.Cs. $\chi(0)=0.21$, $\zeta(0)=-0.4$\\
For $\gamma=-1$: (blue)$\delta=7$,$\kappa=1$-(red)$\delta=5$,$\kappa=5$-(green \& black)$\delta=6$,$\kappa=3$\\
 I.Cs.(blue \& green)$\chi(0)=0.1$, $\zeta(0)=-0.5$
 (red)$\chi(0)=1/3$, $\zeta(0)=-2/9$\\-(black)$\chi(0)=3/10$, $\zeta(0)=-1/3$\\
\end{tabular*}\\

Finally, to validate the model, we check it against observational data from our past light cone such as distance modulus which is defined by $\mu(z)\equiv m(z)-M$, where
$m$ is the apparent magnitude of the source and $M$ its
absolute magnitude. The distance modulus is related to the luminosity distance $d_L$ via $\mu(z)= 5 log_{10} d_L(z) + 25$ where $d_L(z)$ is measured in units of megaparsecs. In Fig 13), the distance modulus obtained from observation at high redshift is compared with the one numerically computed by means of the model using the stability analysis results. From Fig. 13)left) for $\gamma=0$, as can be seen the $\mu(z)$ for three sets of stability parameters $\delta$ and $\delta$  in the model are compared with the observational data. The curves are in general in the best fit with the observational data in the interval $z=0-1$ while for the red curve this continues to the point $z=2$. From Fig. 13)middle) for $\gamma=1/3$, as can be seen the $\mu(z)$ for two stability parameters $\delta=4$ and $\delta=6$  in the model are compared with the observational data. In $\delta=4$ case (blue curve) where the critical point p3 is stable (with $q=-1/7$ and $\omega_{eff}=-9/21$), the graph shows slightly better fit the observational data in comparison with the case $\delta=6$ (black curve) where the model is unstable.

In Fig.13)right) for $\gamma=-1$, as can be seen, the $\mu(z)$ for four different stability parameters are compared with the data. For the stability parameter $\delta=5$, $\kappa=5$ (red curve) which corresponds to an unstable universe, the numerically calculated $\mu(z)$ is out of range. For the stability parameter $\delta=6$, $\kappa=3$ (green curve) where the model contains stable critical points at some stage and the universe passes through the LCDM state for the given initial conditions the model fit the observational data. The black curve has the same stability parameters but with the given initial conditions it does not fit the data. For the stability parameter $\delta=7$, $\kappa=1$ (blue curve) the model contains stable critical points and the universe at some stage passes through LCDM state and as shown the model best fit the observational data.\\

\begin{tabular*}{2.5 cm}{cc}
\includegraphics[scale=.30]{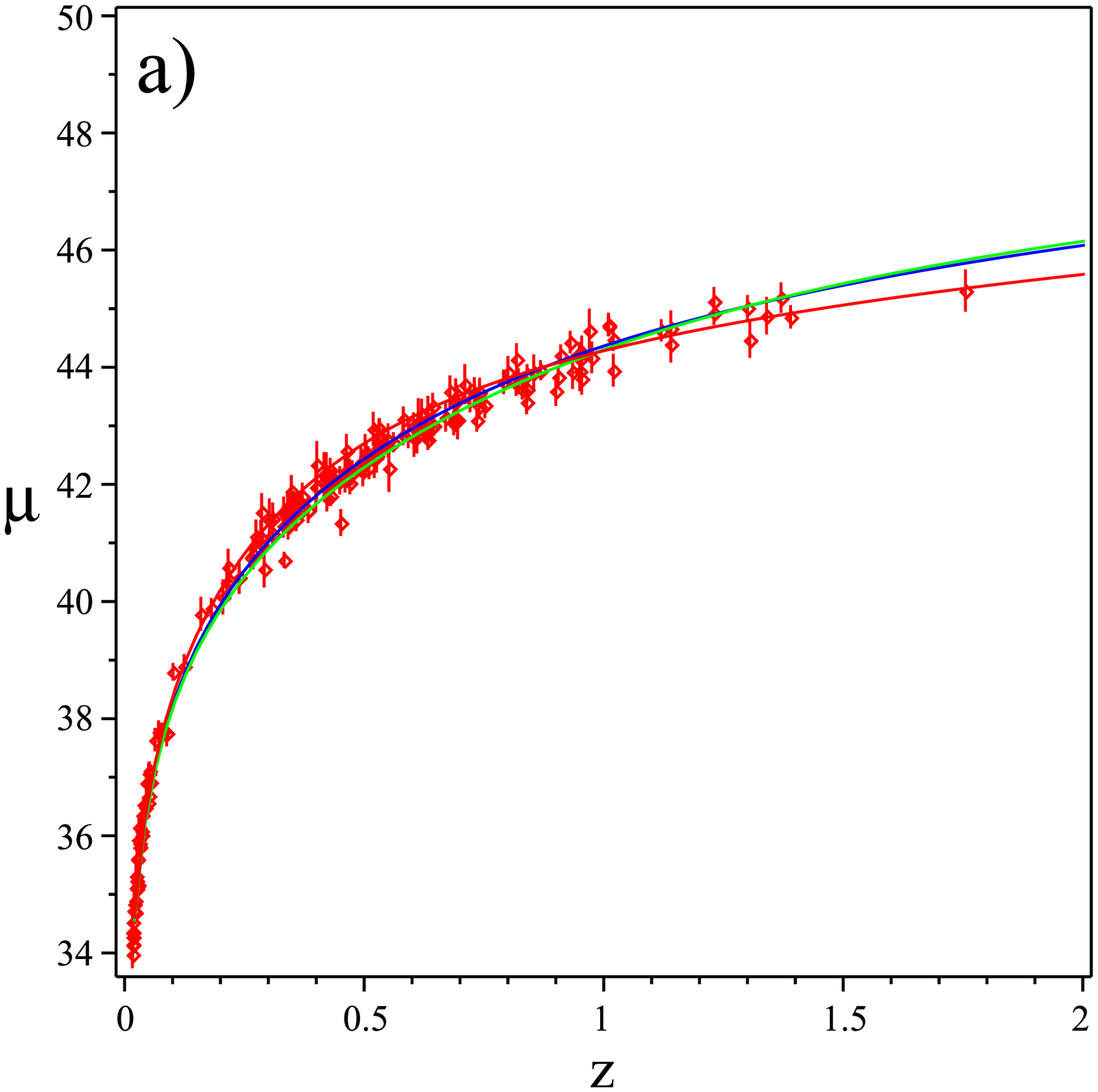}\hspace{0.1 cm}\includegraphics[scale=.30]{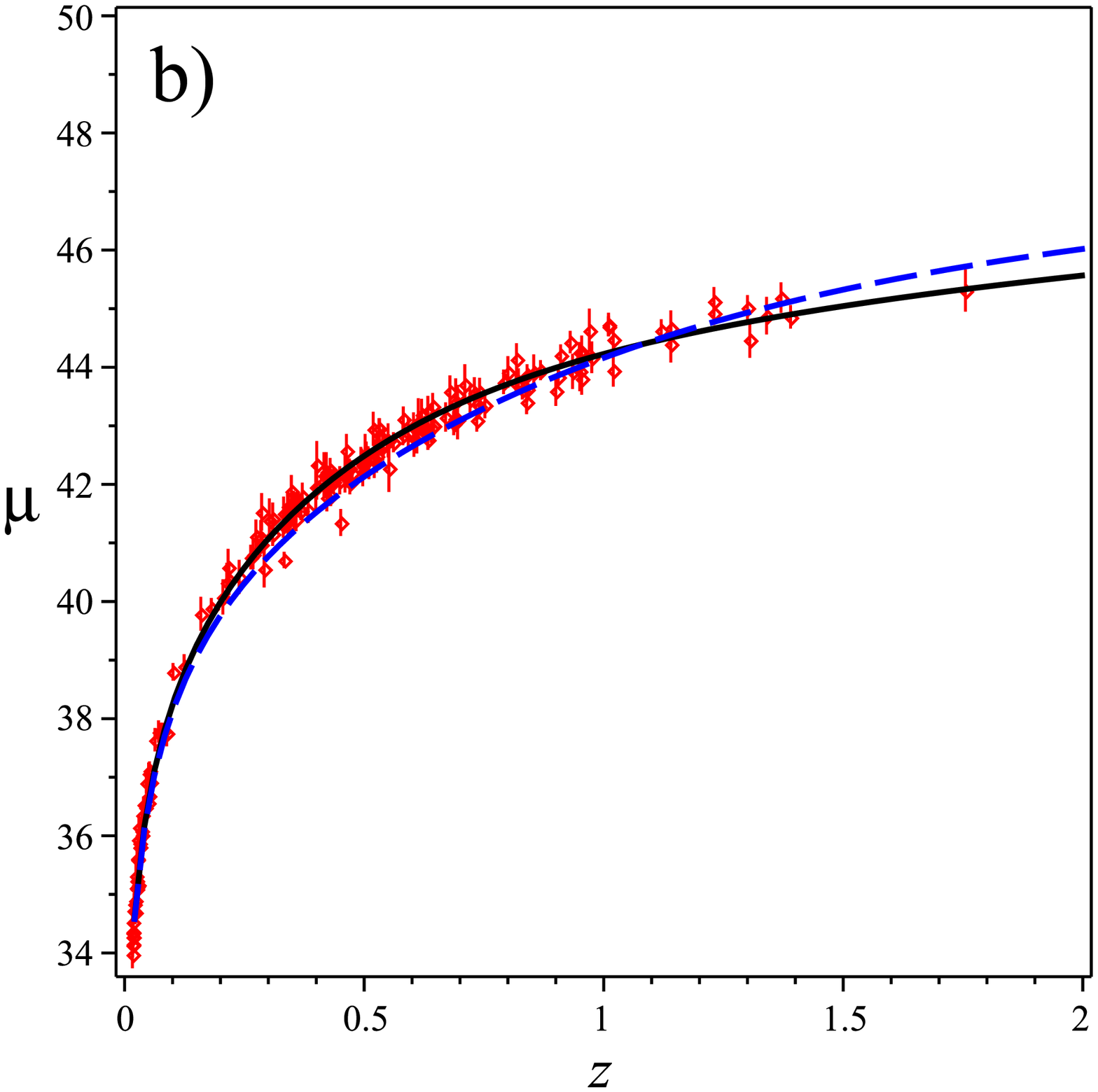}\hspace{0.1 cm}\includegraphics[scale=.30]{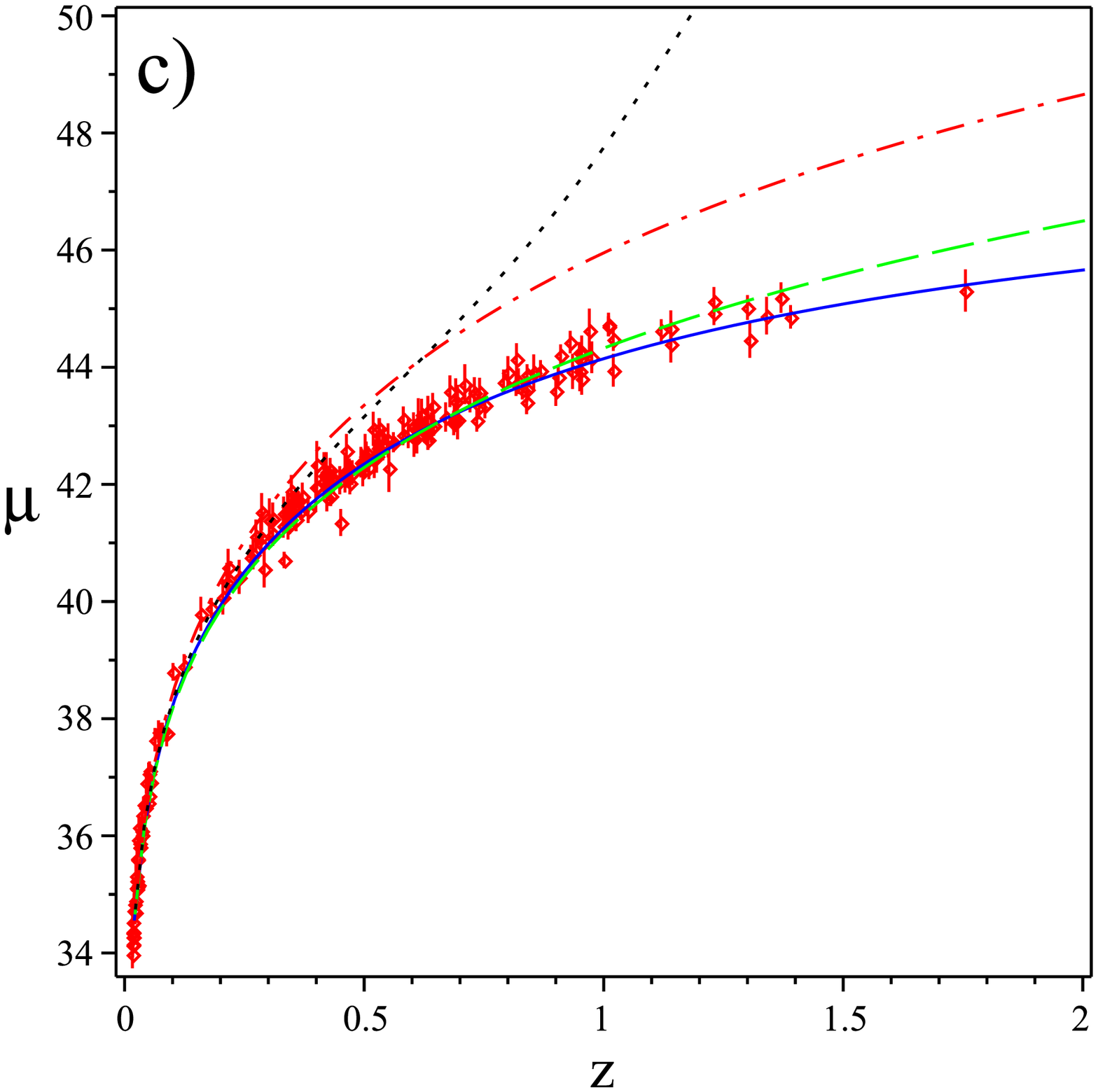}\hspace{0.1 cm}\\
Fig. 13:The graph of the distance modulus $\mu(z0$ for $\gamma=0$(left), $\gamma=\frac{1}{3}$(middle), $\gamma=-1$(right).\\
For $\gamma=0$:(blue)$\delta=7$,$\kappa=1$-(green \&red)$\delta=6$,$\kappa=3$\\
 I.Cs.:(blue \& green)$\chi(0)=0.1$, $\zeta(0)=-0.5$
 (red)$\chi(0)=1/3$, $\zeta(0)=-2/9$. \\For $\gamma=\frac{1}{3}$: (blue)$\delta=4$-(black)$\delta=6$. I.Cs. $\chi(0)=0.21$, $\zeta(0)=-0.4$\\
For $\gamma=-1$: (blue)$\delta=7$,$\kappa=1$-(red)$\delta=5$,$\kappa=5$-(green \& black)$\delta=6$,$\kappa=3$\\
 I.Cs.(blue \& green)$\chi(0)=0.1$, $\zeta(0)=-0.5$
 (red)$\chi(0)=1/3$, $\zeta(0)=-2/9$\\-(black)$\chi(0)=3/10$, $\zeta(0)=-1/3$\\
\end{tabular*}\\

\section{Summery and Conclusion}

This paper is partially designed to study the attractor solutions of the CBD cosmology by
utilizing the 2-dimensional phase space of the theory. The
potential $V(\phi)$ and coupling scalar function $f(\phi)$  in the model is considered to be in power law forms in the phase space.
The matter Lagrangian in the model is regarded as a perfect fluid with three kinds of EoS parameters, $\gamma=0, 1/3,-1$.
The stability analysis gives the corresponding conditions for tracking attractor and determines the universe behavior in the past and future.
By stability analysis, the critical points in the model for the matter fields are evaluated to be stable if satisfying stability conditions, Figs. 2-5.

We then study the cosmological parameters such as effective EoS parameter, $\omega_{eff}$, deceleration parameter, $q$, and statefinder parameters for the model in terms of the dynamical variables introduced in the stability section. The results show that depending on the stability parameters, the model produces the expansion of the universe characteristic of phantom cosmology, despite the absence of phantom energy.  It also shows that the EoS parameter may cross the cosmological divide line depending on the stability conditions in the past and future in all $\gamma=0, 1/3, -1$ cases. Then we perform
a statefinder diagnostic for the CBD model. It is shown that the evolving trajectories of different
scenarios in the $\{r, s\}$ plane is different from those of other DE models. From the graphs we also find that in case of $\gamma=1/3$ the universe may have a cyclic behavior between two unstable states, while passing through LCDM state on its way. 

In conclusion, by fitting the observational data to the model, we compare the distance modulus-redshift relation for $\gamma=0, 1/3, -1$ cases, depending on the stability conditions. In general, in comparison of the observational data with the model for different $\gamma$'s and stability parameters show that the scenarios with phantom crossing better fit the data. Among those scenarios with phantom crossing and best fit with the observational data, the $\gamma=0$ case for the stability parameters $\delta=6$ and $\kappa=3$ ( red curve) is more interesting because the from the Fig.6) we see that the effective EoS parameter, $\omega_{eff}$, and deceleration parameter $q$ start from dust-like dominated era and followed by the dark energy dominated era at the present time. The symmetric behaviour of the graphs shows that the universe continues being in the energy dominated era for a while and finally reaches the dust-like dominated era again.

\end{document}